\documentclass[fleqn,usenatbib]{mnras}

\usepackage{newtxtext,newtxmath}

\usepackage[T1]{fontenc}

\DeclareRobustCommand{\VAN}[3]{#2}
\let\VANthebibliography\thebibliography
\def\thebibliography{\DeclareRobustCommand{\VAN}[3]{##3}\VANthebibliography}

\usepackage{graphicx}	
\usepackage{amsmath}	

\usepackage{xcolor}
\definecolor{arpcolor}{RGB}{51, 166, 255}
\definecolor{tocolor}{RGB}{50, 50, 50}

\title[SF in isolated \& tidally triggered bars]{Differences in star formation activity 
between tidally triggered and isolated bars: a case study of NGC\,4303 and NGC\,3627}

\author[E. J. Iles et al.]{
Elizabeth J. Iles,$^{1}$\thanks{E-mail: iles@astro1.sci.hokudai.ac.jp}
Alex R. Pettitt$^{1,2}$
and Takashi Okamoto$^{1}$
\\
$^{1}$Department of Physics, Faculty of Science, Hokkaido University, Sapporo 060-0810, Japan\\
$^{2}$Department of Physics and Astronomy, California State University, Sacramento, 6000 J Street, Sacramento, CA 95819-6041, USA\\
}

\date{Accepted XXX. Received YYY; in original form ZZZ}

\pubyear{2021}

\begin{document}
\label{firstpage}
\pagerange{\pageref{firstpage}--\pageref{lastpage}}
\maketitle

\begin{abstract}
Galactic bars are important drivers of galactic evolution, and yet how they impact the interstellar medium and correspondingly star formation, remains unclear. We present simulation results for two barred galaxies with different formation mechanisms, bars formed in isolation or via a tidal interaction, to consider the spatially and temporally varying trends of star formation. We focus on the early ($<1$Gyr) epoch of bar formation so that the interaction is clearly identifiable. The nearby NGC\,4303 (isolated) and NGC\,3627 (interaction history) are selected as observational analogues to tailor these simulations. 
Regardless of formation mechanism, both models show similar internal dynamical features, although the interaction appears to promote bar-arm disconnection in the outer disc velocity structure. Both bars trigger similar boosts in star formation (79\%; 66\%), while the interaction also triggers an earlier 31\% burst. Significant morphological dependence is observed in the relation between surface gas and star formation rate. In both cases, the bar component is notably steepest; the arm is similar to the overall disc average; and the inter-arm clearly the shallowest. A distinguishable feature of the tidal disc is the presence of moderately dense, inefficiently star forming gas mostly confined to tidal debris outside the optical disc. The tidal disc also exhibits a unique trend of radially increasing star formation efficiency and a clear dearth of star formation which persists along the bar between the centre and bar ends. These are potential signatures for identifying a barred system post-interaction. 
\end{abstract}





\section{Introduction}
\label{s:intro}
Understanding the formation and evolution of galaxies is important for determining the properties and processes that govern physics in the Universe. In modern astrophysics, barred-spiral galaxies in particular are of prime interest due in part to the likely similarity in morphology with the Milky Way. In addition, such galaxies have been shown to make up a large portion of all observed spiral galaxies, although the exact fraction ($\sim25$--$75$\%) depends heavily on bar classification criteria \citep{Schinnerer2002,Aguerri2009, Masters2011}. It has been shown that a range of formation mechanisms can produce such a bar-like feature in the central region of galaxies. For example, it is possible for bars to form in kinematically cold, sufficiently massive, isolated stellar discs \citep{Hohl1971,Ostriker1973}. Under this regime, an initial instability drives a fast growth phase where the bar emerges and subsequently buckles out of the disc-plane, before slowing down and growing gradually in a phase of secular evolution \citep{ Raha1991,Sellwood2014}. 

Under the $\Lambda_{\rm CDM}$ cosmology, wherein structure forms hierarchically, morphology can be influenced by the associated galaxy-galaxy interaction \citep{Miwa1998,Lang2014,Romano-Diaz2008,Pettitt2018}. Historically, it was shown that these interactions can induce bar formation in isolated, bar-free galaxies or slow the formation of a bar in an isolated disc galaxy that was already likely to form a bar; although, this was likely to have negligible effect on the properties of the bar itself \citep{Noguchi1987,Salo1991}. Conversely, subsequent studies have proved that some interaction conditions can instead dampen or completely halt bar formation, rather than induce it \citep{Athanassoula2002,Kyziropoulos2016,Gajda2017,Moetazedian2017,Zana2019}; and that bars which form in interacting systems instead appear to rotate slower than those formed in isolation \citep{Miwa1998,Martinez-Valpuesta2017,lokas2018}. It is also argued that the presence of a bar must have some environmental dependence, which could imply a subsequent co-dependence for bar formation on interaction \citep{Mendez-Abreu2012,Skibba2012,Pettitt2018}. 

Regardless of which factors influence the formation of the central bar, it is widely accepted that the presence of such a feature has significant effect on a range of properties within the host galaxy \citep{Sheth2002,Fujii2018,Zana2019}. For example, star formation rates for regions within morphological features, such as the bar and arms, are consistently observed to differ considerably, often showing significantly lower star forming efficiency in the bar compared to other areas of the disc \citep{Downes1996,Sheth2002,Watanabe2019}. This variation is often attributed to the non-circular motions of stars and gas in the bar, usually induced by a non-axisymmetric bar potential that generates strong shock or shear motions along the bar and disrupts the bar-located molecular clouds, thus suppressing star formation in that region \citep{Roberts1979,Athanassoula1992,Schinnerer2002,Dobbs2014,Beuther2018}. As such, barred-spiral galaxies are often considered when probing the relationship between star formation on small molecular cloud scales (10--100\,pc) and galactic dynamics or gas kinematics at $\sim$kpc scales \citep{Kuno2000,Warren2010,Watanabe,Hirota2014}.

While the presence of a bar clearly affects the star formation properties within a single galaxy, interactions between galaxies have also been shown to strongly affect the gas dynamics and gas fractions of each involved galaxy. This will consequently also affect the star formation properties of these galaxies. One such effect is the so-called starbursts -- accelerated star formation triggered by the inflowing gas provided through the interaction forces \citep{Mihos1994,Barnes1996,Hopkins2008}. These driven starburst events are mostly observed in the central region of the galaxy; however, there is evidence to suggest that this effect is not limited to the central region but also observed well into the disc of the galaxy, or even only in the disc in some cases \citep{DiMatteo2007,Bournaud2011,Pettitt2017}. Interactions in the form of minor mergers, also appear to significantly impact star formation statistics at low redshift \citep{Darg2010,Kaviraj2014,Taylor2017}. Such interaction driven effects on the star formation properties of galaxies are not necessarily distinguishable from similar effects which can be attributed to bars in isolated, unperturbed galaxies. Hence, tracers of the mechanisms which drive and impact bar formation, such as tidally driven bars, may not be entirely separable from those galactic properties which are influenced by the presence of the bar itself.

Increasingly detailed, high resolution, large-scale observational surveys continue to improve our capability to analyse star formation rates and efficiencies, as well as document the wide range of observed morphological features in galaxies --i.e. the COMING--CO multi-line imaging of nearby galaxies \citep{Sorai2019}, PHANGS---Physics at High Angular resolution in Nearby GalaxieS \citep{Lee2021} survey results utilised in the following sections. However, regardless of the breadth and depth of modern observational capabilities, it is simply impossible for observational surveys to capture long term evolution of an individual galaxy. 

High resolution numerical simulations have long been identified as important in this area, as these provide the capability to track changes in properties over long time-scales and at varying length scales. The formative work of \citet{Toomre1972a} and their contemporaries have cemented N-body simulations as an indisputably effective tool for determining precise mechanisms and kinematics which drive the formation of structure in observed galaxies. Since then, due to constant scientific and technological advancement, it has become possible to produce simulations which can even replicate the evolution of a galaxy under self-gravitation at a resolution comparable to a true number of stars \citep{Fujii2018}. The mechanisms which may affect bar and arm features have been studied extensively in this manner, including but not limited to internal properties of individual galaxies, such as mass-fractions and internal forces \citep{Hohl1971,Ostriker1973,Friedli1993,Baba2015,Wu2015,Sellwood2020}, and all manner of interactions constraining size-ratios, orbits and internal structures \citep{Noguchi1987,Elmegreen1991,Fiacconi2012,Dobbs2013,Oh2015,Pettitt2018}. 

Numerical methods are additionally indispensable in the effort to completely resolve star formation traits and properties of galaxies \citep{DiMatteo2007, Springel2010, Federrath2012, Hopkins2013, Okamoto2015}. The advent of detailed numerical studies of galactic discs has made significant advancements to the global understanding of the impact that specific disc features, such as the bar and spiral arms, may have on star formation processes at a range of scales from molecular clouds to region averaged star formation rates and efficiencies \citep{Bournaud2010, Renaud2013, Cole2014, Fujimoto2016, Baba2020}. For example, through simulation, shear in different disc environments has been shown to affect cloud formation conditions and cloud-cloud collisions, producing differing likelihoods for the formation of massive stars or star formation efficiency in different morphological features, such as the bar ends or feathered and spurred offsets to the arms \citep{Dobbs2006, Emsellem2014, Renaud2015, Takahira2018}. Additionally, it has been possible to explore the limits of the so-called Kennicutt-Schmidt relation \citep{Schmidt1959, Kennicutt1987} in terms of both very high spatial resolution over a complete galactic disc as well as morphologically distinct features such as bars and arms \citep{Fujimoto2014}. However, in most previous numerical studies of star formation in disc galaxies, the focus has been predominantly on either the differing affects inherent to various observed morphological features \citep{Bournaud2010,Fujimoto2014,Renaud2015,Baba2020} or the influences of the external drivers governing the formation of such features \citep{Tan2000, DiMatteo2007, Inoue2013, Pettitt2018}. Few have attempted to look at whether different origins for morphological features, such as the galactic bars studied herein, may impact the affect of these features on the star-forming ISM. 

In this work, we employ numerical hydrodynamics to produce simulations for two specific case studies: a bar formed in isolation by random perturbation and a bar triggered by the tidal forces of an interaction. These case studies are determined to be consistent with real, observed target galaxies: NGC\,4303 (M61), noting it has been determined to be an isolated disc galaxy on the outskirts of the Virgo cluster with no HI gas depletion \citep{Yajima2019}; and NGC\,3627, a counterpart determined to have obvious traces of past interaction through asymmetry and distorted HI gas \citep{Haynes1979}. A brief summary of the observationally determined properties of these targets is presented in Section \ref{s:targets}. Section \ref{s:methods} outlines the simulation initial conditions and methods. The relevant findings are presented in Section \ref{s:results}, specifically those pertaining to general galaxy attributes, star formation features and the evolution of morphologically dependent properties. This is accompanied by a discussion of the relevant scientific implications for such results in Section \ref{s:discussion}. Conclusions and avenues for future work are summarised in Section \ref{s:conc}. 

\section{Targets}
\label{s:targets}
These target galaxies were selected primarily due to the wealth of available observational survey data and results, both previous and ongoing. They have long been determined to be of barred-spiral type with relatively face-on inclinations and local distances, conducive for taking relatively detailed observational measurements. Thus, observations of these nearby galaxies were used to situate the following simulations into a context which is physically relevant. 

NGC\,4303 is a nearby barred-spiral galaxy which is likely to be associated with the Virgo cluster of galaxies \citep{Binggeli1985,Ferrarese1996}. Due to the relatively straightforward geometry (almost completely face-on to the line-of-sight), the active galactic nuclei of LINER/Seyfert 2 type \citep{Ho1997}, and potential double bar, NGC\,4303 has been the target of a number of observational studies, particularly focused on the central region and constraining its dynamics or morphology \citep{Sofue1997,Helfer2003, Kuno2007,Momose2010}. It has been determined to be an Sbc-AB type galaxy with an approximate distance of 16.1--17.6 Mpc with an inclination angle of 25.0 degrees \citep{Schinnerer2002,Utomo2018}. The observationally derived rotation curve is generally flat with deviation only in the very central region where it peaks sharply \citep{Guhathakurta1988, Sofue1997,Lang2020}. \citet{Egusa2009} attempted to determine a pattern speed, $\Omega_P$ , from CO--H$\alpha$ offsets but, due to insufficient measurements, the uncertainty dominates ($\Omega_P \sim 24 \pm 29$\,km\,s$^{-1}$\,kpc$^{-1}$). Other studies, both through observation and simulation, have also sought to derive pattern speeds for the central bar region, although these are considerably inconsistent based on the determination of what radius constitutes this region, ranging from $R_{\rm centre} = 2.8 \sim 9.2$\,kpc \citep{Colina2000,Rautiainen2005,Egusa2009}.

\citet{Utomo2018} derive the stellar mass as $\sim7.943 \times 10^{10}$\,M$_\odot$ and SFR of 5.248\,M$_\odot$\,yr$^{-1}$ from CO measurements at 120\,pc resolution. NGC4303 also has an estimated average time for massive star formation from molecular clouds in the spiral arms ($t_{\rm SF}$) of 10.8 $\pm$ 5.7\,Myr \citep{Egusa2009} which is consistent with the age of young clusters in H$\alpha$ determined by \citet{Koda2006} to be approximately 10\,Myr. The gas mass derived was $5.3 \times 10^9$\,M$_\odot$ with average surface density of 36\,M$_\odot$\,pc$^{-2}$ across $160^{\prime\prime}$ ($\sim12$\,kpc) in the molecular disc \citep{Momose2010}. From this, NGC\,4303 is determined to have an overall average SFR surface density of $8.3 \times 10^{-2}$\,M$_\odot$\,yr$^{-1}$\,pc$^{-2}$ with the bar contributing $7.6 \times 10^{-2}$\,M$_\odot$\,yr$^{-1}$\,pc$^{-2}$, approximately 10\% lower than the overall disc average, and arms $9.8 \times 10^{-2}$\,M$_\odot$\,yr$^{-1}$\,pc$^{-2}$, approximately 10\% higher than the overall disc average \citep{Momose2010}. Subsequent studies confirm this observation that NGC\,4303 shows a lower SFE in the bar than in the spiral arms \citep{Yajima2019,Muraoka2019}. In general, when compared to the distribution of galaxies assessed by \citet{Kennicutt1998}, NGC\,4303 appears to have higher than average star formation activity (factor of $\sim5$) according to observations \citep{Momose2010,Yajima2019}.

Comparatively, NGC\,3627 is a barred-spiral galaxy of type SABb according to the Third Reference Catalogue of Bright Galaxies \citep{DeVau1991}. Like NGC\,4303, it has an active nucleus of the LINER/Seyfert 2 type AGN \citep{Ho1997}. It is widely considered to have experienced some interaction with nearby NGC\,3623 and NGC\,3628 at some point in the evolutionary history of both galaxies due in particular to the slight asymmetry in arm structure observable at optical wavelengths and a distorted HI gas distribution \citep{Haynes1979}. Additionally, significant distortion is also evident in the disc in polarised maps soft X-ray emission (0.2--1\,keV), as well as a large asymmetry in hot gas temperature on the bar edges which implies a recent collision with a dwarf companion galaxy \citep{Wezgowiec2012}. The inclination angle of approximately 60 degrees has made NGC\,3627 a relatively popular target for CO mapping and the study of molecular gas properties, as well as star formation activity \citep{Reuter1996,Helfer2003,Kuno2007}. \citet{Hirota2009} determined it to have pattern speed of $\Omega_P = 39$\,kms$^{-1}$kpc$^{-1}$ from measurements of $^{12}$CO (1-0) data and the method prescribed by \citet{Kuno2000}. \citet{Law2018} estimate NGC\,3627 to have a total dynamical mass of $4.94 \pm 0.7 \times 10^{10}$\,M$_\odot$ at a galactocentric radius of $\sim6.2$\,kpc (121$^{\prime\prime}$) from CO (2-1) emission measurements.

Additionally, there is also postulated to be a correlation between the kinetic temperature of gas in NGC\,3627 and star formation efficiency \citep{Law2018}. Considering the relationship between molecular gas and star formation activity, NGC\,3627 shows significant differences in the SFR from region to region but the bar itself has very low observable star formation efficiency \citep{Watanabe}. The bar ends, however, show the most intense star formation of any region with SFE for the three regions Spiral Arm, Bar-End, and the Nuclear Region determined by \citet{Watanabe2019} to be $1.3\pm0.4$, $5.7\pm1.7$ and $1.8\pm1.0 \times 10^{-9}$\,yr$^{-1}$ respectively. Despite this, the chemical composition in these regions appears to be quite similar, indicating that the characteristic chemistry on the observed scale is generally insensitive to physical conditions and local effects, such as the star formation rate \citep{Watanabe2019}. Many studies of NGC\,3627 are concerned by this significant variation between regions, particularly at the bar ends and have endeavoured to attribute physical effects to such results \citep{Watanabe,Casasola2011,Law2018,Beuther2018,Watanabe2019}. 

\begin{figure}
	\includegraphics[width=\columnwidth]{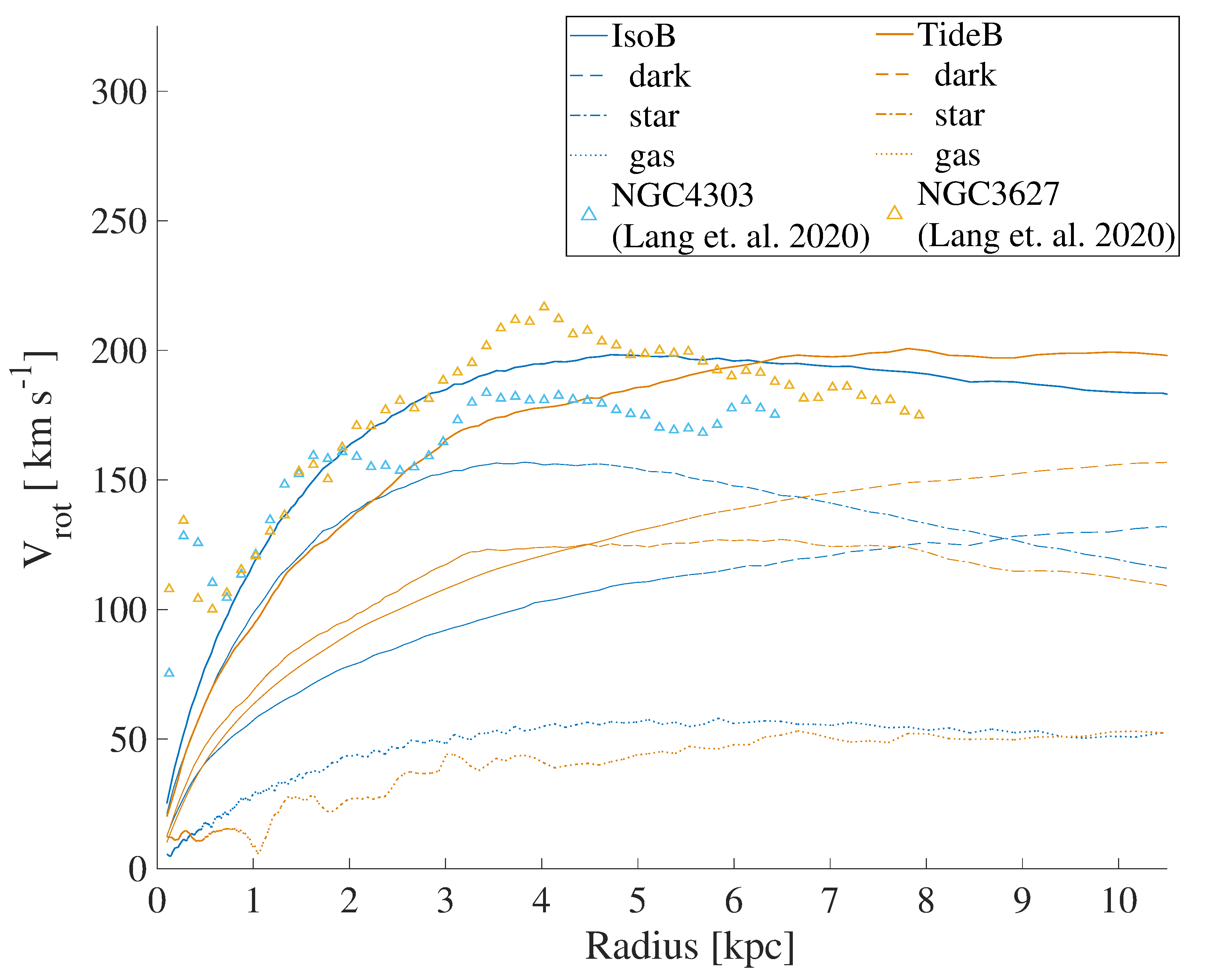}
    \caption{Rotation curves from initial conditions set for each case. IsoB is identified by blue lines and TideB by orange. The different component contributions are denoted by various line styles (solid--total, dashed--dark, dot-dashed--stellar, dotted--gas). An example of observational measurements for the two targets used to constrain appropriate initial conditions is also represented by triangle points from the PHANGS survey data \citep{Lang2020}.}
    \label{f:rc_compare}
\end{figure}

\section{Methods}
\label{s:methods} 
\subsection{Initial Conditions}
To compare the effect of different formation mechanisms on the properties of barred-spiral galaxies, initial conditions which would evolve to form a bar feature were developed for the two case studies: the first, which relied on an initial instability within the disc to trigger bar formation; and the second, which instead required the external impetus of a passing companion. These initial conditions were also constrained to specifically align with surface density profiles and kinematic data of the observed target galaxies (NGC\,4303, NGC\,3627).

To develop an isolated disc with a bar (hereafter IsoB), the \textsc{galic} package of \citet{Yurin2014} was used to produce an initial condition which would evolve into a barred-spiral galaxy without external influence. Defining initial parameters with \textsc{galic}, a spherical dark halo and stellar bulge were characterised by a Hernquist profile with an axisymmetric velocity structure of distribution function $f(E,L_z)$ and a specified net rotation value. A stellar disc component was defined to be a thin disc of thickness 0.2 times the radial scale length. An axisymmetric velocity structure was also enforced in the disc with a distribution function of the form $f(E,Lz,I_3)$ and a set disc dispersion ratio between the radial and vertical velocities ($\langle v_z\rangle ^2 / \langle v_r\rangle ^2$) was enforced, as well as a net rotation value. Within these general descriptions, it was possible to refine an initial condition which produced a disc galaxy physically comparable to the desired target by simply modifying initial stellar mass ratios ($m{_{\rm disc}}, m{_{\rm bulge}}$) and the disc velocity dispersion. Through an iteration of test initial conditions evolved at low resolution, these parameters were selected to be $m{_{\rm disc}} = 0.065, m{_{\rm bulge}} = 0.01$, as a fraction of total mass; and a disc dispersion ratio of $\langle v_z\rangle ^2 / \langle v_r\rangle ^2 = 1.5$. 

\begin{table}
	\centering
	\caption{Initial component masses and scale lengths for the two models IsoB and TideB. Masses are given in units of $10^{10}$\,M$_\odot$ while distances are reported in\,kpc. The distance parameter for the companion is defined by the closest approach ($b$).}
	\label{t:IC}
	\begin{tabular}{lccccccccc} 
		\hline
		&\,$M_{_{\rm gas}}$ &\,$M_{*_{\rm disc}}$ &\,$M_{*_{\rm bulge}}$ &\,$M_{_{\rm halo}}$ &\,$M_{_{\rm companion}}$ \\
		\hline
		\textbf{IsoB} & 0.522 & 2.611 & 0.402 & 37.16 & - \\
		\textbf{TideB} & 0.759 & 2.441 & 0.072 & 42.57 & 2.401\\
		\hline
		& $a_{_{\rm gas}}$ & $a_{*_{\rm disc}}$ & $a_{*_{\rm bulge}}$ & $a_{_{\rm halo}}$ & $b_{_{\rm companion}}$ \\
		\hline
		\textbf{IsoB} & 3.090 & 2.060 & 2.057 & 20.57 & -\\
		\textbf{TideB} & 3.705 & 2.470 & 0.405 & 20.26 & 10 \\
		\hline
	\end{tabular}
\end{table}

The appropriate parameters for these tests were selected by matching the presence of morphological features, such as the presence of a distinguishable bar, with an elementary assessment of the galaxy rotation properties. The rotation curve for a range of appropriate initial conditions was compared with observed rotation curves for the target isolated galaxy NGC\,4303 \citep{Guhathakurta1988,Sofue1997,Yajima2019,Lang2020}. The rotation curve of the initial conditions used for each final case is shown in Figure \ref{f:rc_compare}. These are decomposed by component and are shown alongside an example of the observationally measured values of \citet{Lang2020}. In both cases, as this figure shows only the initial condition rotation curves, these profiles are generally flat, dominated by the stellar component within the disc radius and dark component outside of this. The overall shape is considered consistent with the observation at first order before developing the small scale features, such as the peaks and wiggles seen in real galaxies as they evolve. The minimum resolution to ascertain the quality of these initial conditions was set at a threshold of $10^6$ disc particles. This limit has been shown to be a requirement in order to properly capture the spiral and bar features in similar N-body galaxy simulations \citep{Fujii2011}.

For the purpose of studying star formation, a gas component is also required in the initial condition, however this is not included with \textsc{galic} determination. To introduce a gas disc, additional particles were subsequently added. This component was comprised of the existing stellar disc particles produced by \textsc{galic}, rotated in the disc plane 180 degrees out of alignment with their stellar counterparts and expanded to 1.5 times the stellar radius. The gas particle mass was also modified to be comparable with the observed gas-mass fraction and derived values for the total gas mass of the target galaxy from observations. This component is singular and does not differentiate between different kinds of gas contributions, for example, molecular or atomic gas components. The complete initial condition for the isolated disc case can be described in terms of a combined array of mass and scale length parameters which are listed in Table \ref{t:IC}. With $5 \times 10^6$ gas particles, the mass resolution of this model is 1044\,M$_\odot$. This resolution is sufficient to assess the coarse-scale properties of GMCs, though as we lack a treatment of molecular chemistry we defer such analysis to future works with a more detailed stellar physics and chemical prescription. An assessment of the likelihood for this disc to form a bar was also performed in the form of \citet{Efstathiou1982} where;
\begin{equation}
\epsilon_{\mbox{bar}} = \frac{V_{\mbox{max}}}{\sqrt{G M_d/a_d}}
\end{equation}
Here, $G$ is the gravitational constant and $V_{\mbox{max}}$ the maximum rotation velocity. The IsoB disc produces a value of $\epsilon_{\mbox{bar}} \sim 0.8$ which is an indication that this disc will be distinctly unstable to bar formation. 

For the second case, a bar triggered by a tidal interaction (hereafter TideB), we use a modified version of the Rise-S10 simulation of \citet{Pettitt2018}. While \citet{Pettitt2018} study a number of tidal interactions, Rise-S10 is the clearest case of a bar being tidally induced (e.g. the Mid and Fall models tended to still form bars in isolation, the interaction merely modified the bar structures). The original Rise-S10 initial condition was appropriately rescaled in both mass and scale length to be comparable to both the isolated initial condition and observational values for the interacting target. Briefly, this corresponds to scaling by factors 0.5, 0.4 and 2 for distances, stellar mass and gas mass respectively. Similar to the IsoB case, this initial condition was constrained to align with surface density profiles of the observed target galaxy. The specific orbit of the companion is not intentionally constrained to any specific observational values of the target galaxy or its neighbours. In this case, the orbit is as per the original system in \citet{Pettitt2018} and is defined purely to trigger an interaction which will drive the desired bar formation. The TideB initial rotation curve decomposed into components can also be seen in Figure \ref{f:rc_compare}. Similarly, the rotation curve is generally flat but is notably less dominated by stellar mass in the central region than the IsoB set of conditions. The maximum rotation speed is also marginally higher between the two cases. As this is the initial condition rotation curve, it is as yet unperturbed by the companion and so is expected to differ more significantly from the observed rotation curve particularly in the central region. After the interaction, this curve undergoes a boost in rotation speed, particularly in the central region. The specific evolution of the rotation curve for both cases is presented in Appendix\;\ref{a:rc}. 

The same mass and scale length parameters are used to describe the initial conditions for the tidally affected disc, including the companion (see Table \ref{t:IC}). In this case, the gas mass resolution is 1084\,M$_\odot$ which is comparable to the isolated model ($\sim$ 1000\,M$_\odot$). The \citet{Efstathiou1982} metric for bar stability was also calculated for this disc without influence from the companion to produce an $\epsilon_{_{\mbox{bar}}} \sim 1.0$ which is significantly greater than the isolated disc, indicating that this model would otherwise be relatively more stable against bar formation. Thus, two initial conditions are produced corresponding to an unstable isolated disc and a tidally driven disc which should each generate a morphologically distinct central bar feature.

\subsection{Simulation Parameters}

\begin{figure*}
	\includegraphics[width=\textwidth]{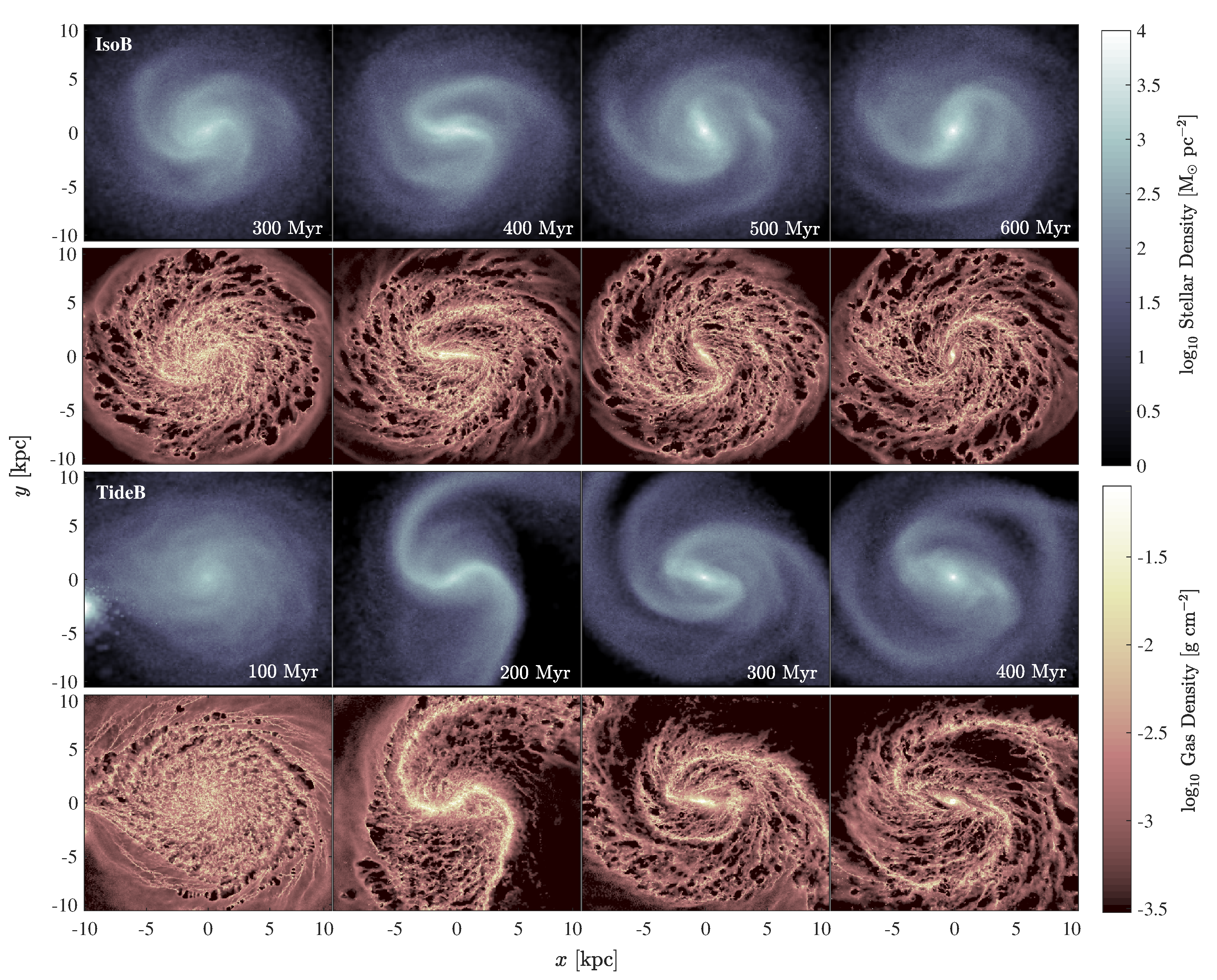}
    \caption{A face-on representation of stellar (top) and gas (bottom) populations, setting the disc plane into the $xy$-plane for both the IsoB and TideB cases. disc rotation is counter-clockwise and the columns step in time by 100\,Myr (one time pre-bar and three post bar formation) showing morphological evolution. The companion can also be seen in passing at the far left of the first TideB window in the sterllar map (100\,Myr). }
    \label{f:stargas}
\end{figure*}

The subsequent simulation process was performed using the \textsc{gasoline2} smoothed particle hydrodynamics (SPH) code \citep{Wadsley2004,Wadsley2017}. We use the standard hydrodynamical treatment advocated by \citet{Wadsley2017}: using 200 SPH nearest neighbours and the Wendland C4 kernel \citep{Dehnen2012}. Gravitational softening lengths of $[0.1, 0.05, 0.01]$\,kpc were used for halo, stars and gas respectively in both simulations. The star formation conditions were prescribed by upper temperature and density thresholds of 100\,atoms/cc (minimum density for forming stars)  and 300\,K (temperature limit for forming stars), respectively. The star forming efficiency was defined by a value of $C_* = 0.1$ ($\sim10\%$ efficiency) following the standard \textsc{gasoline2} sub-grid prescriptions \citep{Katz1996,Stinson2006}; with a \citet{Chabrier2003} IMF and a convergent flow requirement for star formation. These values are similar to previous studies \citep{Saitoh2008, Tasker2008,Pettitt2017}. The effects of UV and photoelectric heating, as well as metal cooling were accounted for with a tabulated cooling function \citep{Shen2010}. From an initial isothermal gas ($10^4$ K), this cooling produces a two-phase thermal profile comparable to the ISM \citep{Wolfire2003}. Stellar feedback is implemented using the \citet{Keller2014} supernova super-bubble feedback with $10\%$ feedback efficiency. This feedback is generated from clusters of young stars instead of individual supernovae and appears more efficient at describing gas motion, regulating star formation and producing the expected strong outflows \citep{Keller2014}. Both models were integrated until it was possible to identify a clear bar feature in each case. The integration was then allowed to progress for a further period of $\sim200$\,Myr (approximately another full galactic rotation) before undergoing the analysis presented in the following sections. This corresponds to a period of evolution where the effects of the tidal perturbation are most prominent in the interacting case, thus highlighting any fundamental differences which arise and may be particularly attributed to the variation between these two types of bar formation mechanisms.

\section{Results}
\label{s:results}

\subsection{General Results}
\label{ss:gen_results} 

\subsubsection{Morphological Structures}
We first present the general morphological features of our two models for bar formation. Figure \ref{f:stargas} shows a face-on view of the stellar and gas populations with the disc plane set onto the $xy$-plane for both the IsoB and TideB cases. The colour weighting is held constant over both cases. In this figure, the top row of each case corresponds to the stellar component and the second row to the gas component. The columns correspond to steps of 100\,Myr, commencing one period immediately preceding a clearly identifiable bar forming. Both cases produce an essentially two-armed structure with a central bar, although there are obvious visual differences within this general morphology. 

The IsoB case, which formed a bar through some initial disc instability, seems to evolve from a more flocculent type galaxy into a two-armed structure as the bar grows. Conversely, the TideB case is strongly affected by the interaction occurring at approximately 100\,Myr (closest approach $\sim94$\,Myr), producing clear, crisp two-arm features. These features however, are significantly impacted by gravitational effects, both from the inner disc and the external companion, constantly strengthening and decoupling over the period of interest. As the simulation evolves, the bar forms and the disc stabilises. Due to the decoupling of the arms in the tidally driven case the bar length intermittently appears to extend. This may also affect an assessment of bar orientation. 

Additionally, it is possible to see that the tidally driven case produces a pronounced bar feature significantly faster than the isolated disc. For convenience, we present all subsequent results in terms of a shifted time scale ($t^\prime$), relative to the bar formation time (rounded to the nearest 100\,Myr) such that $t^\prime = 0$ at the approximate period of distinct bar formation. This revised time of $t^\prime = 0$ corresponds to the $t = 400$\,Myr snapshot of IsoB and the $t = 200$\,Myr snapshot for TideB which can be seen in Figure \ref{f:stargas}. At a given evolutionary period (within each column of Figure \ref{f:stargas}) when the bar is well developed, the large-scale structural features of each case do not appear substantially different. The bar lengths and strength appear visually similar within each disc. The number and prominence of the arms, as well as their pitch angles, are also similar.

\begin{figure}
	\includegraphics[width=\columnwidth]{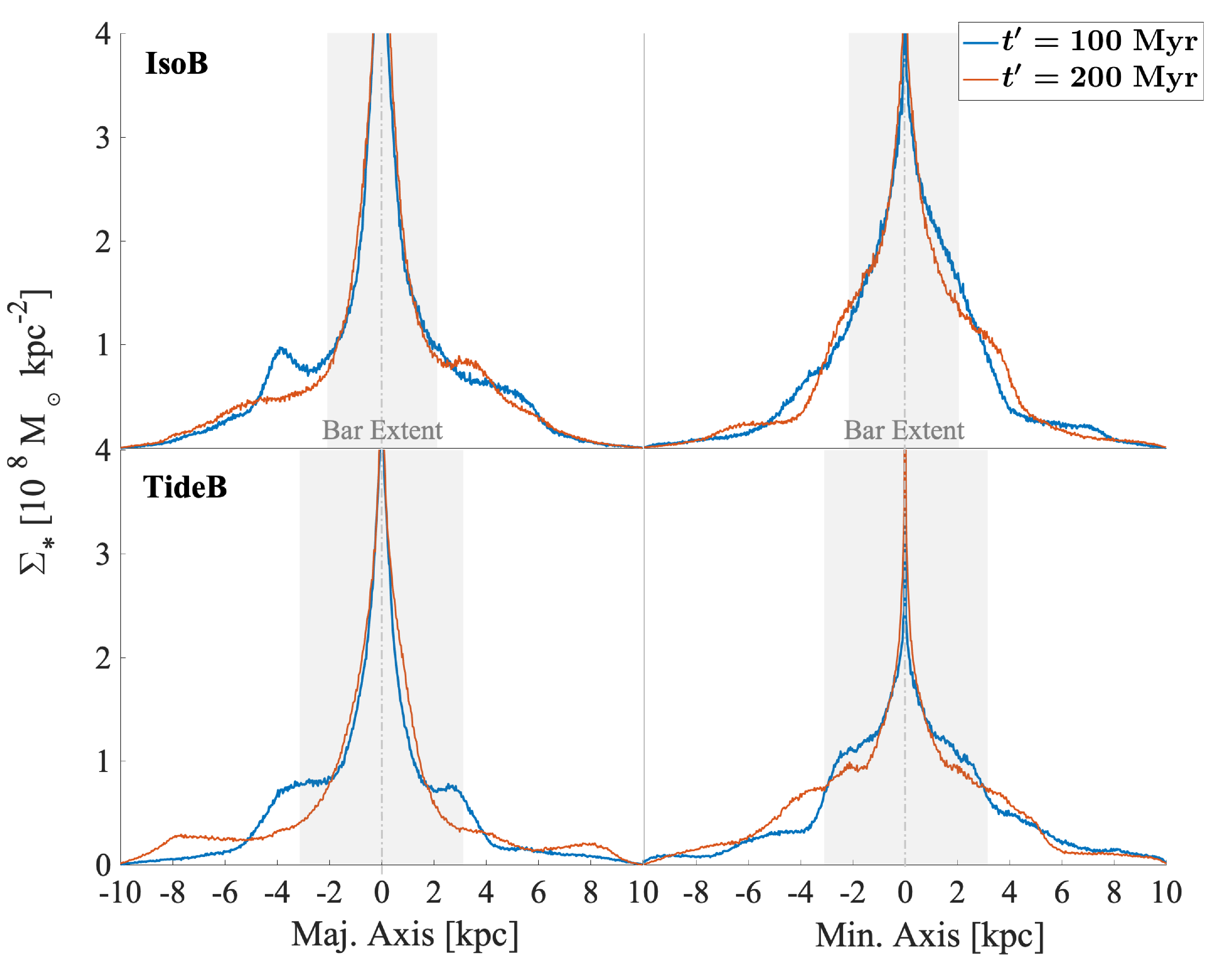}
    \caption{Stellar surface density profiles along the major (left) and minor (right) axis of the bar for each case in the time periods where the bar is well defined ($t^\prime = 100, 200$\,Myr). The region shaded in grey indicates the bar extent calculated as in Section \ref{sss:sf_history} (IsoB: $R_{\rm bar}=2.10$\,kpc; TideB: $R_{\rm bar}=3.12$\,kpc).}
    \label{f:bardensity}
\end{figure}

Figure \ref{f:bardensity} shows the stellar surface density profile for each case, both along the bar (major axis) and perpendicular to it (minor axis). 
Each model exhibits very similar profiles both parallel and perpendicular to the bar orientation. Stellar surface densities within the central region clearly decrease exponentially and in a manner which is noticeably steeper than the surrounding regions that are solely composed of the non-barred portion of the disc. This sharply exponential shape is noticeably prevalent in a much narrower section of the bar minor axis as would be expected from the elliptical shape of the bar morphology that can be seen in Figure \ref{f:stargas}. 

Both cases can be categorically described as hosting very similar bars of exponential (late) type, rather than the alternative flat (early) type \citep{Elmegreen1985}. Both NGC\,4303 and NGC\,3627 have been determined to have later type Hubble classifications \citep{DeVau1991}. Hence, this is in line with expectations for the target galaxies as well as the exponential disc profile used to produce the initial simulation conditions. 
It may thus be naively expected that these two bars should generally impact the disc structure in a similar manner.

\subsubsection{Velocity Structure}
Non-axisymmetric morphological features are known to act as perturbations to the galactic velocity field (e.g. \citealt{deblok2008}). The two target systems in particular are known to show clear non-circular motions, both from moment maps and undulations in the rotation curves \citep{Law2018,Schinnerer2002}.

\begin{figure*}
	\includegraphics[width=\textwidth]{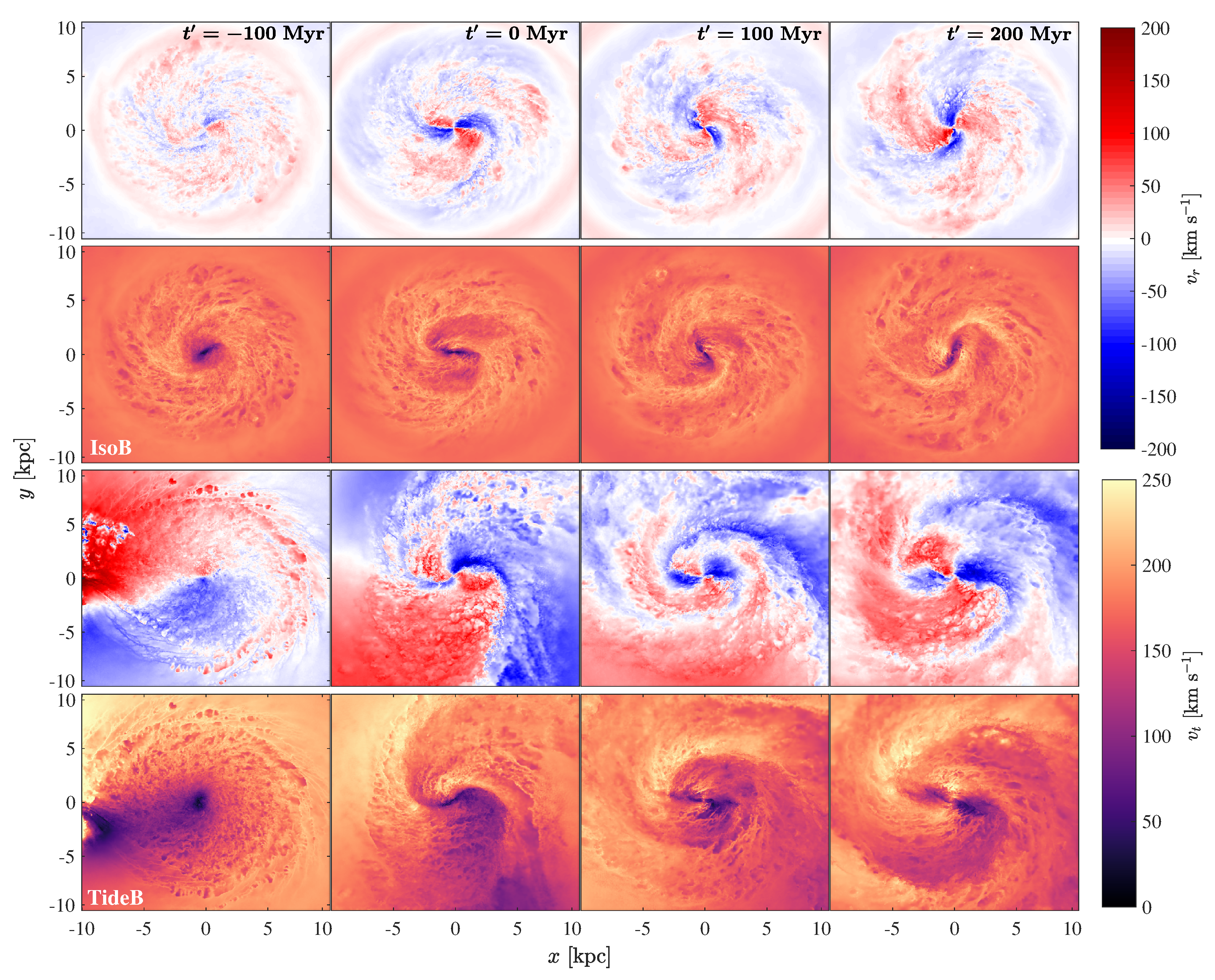}
    \caption{The velocity of gas projected into the same $xy$- (disc-aligned) plane. For each case, radial (top) and tangential (bottom) velocities are shown. The colour scale is set to be consistent over both cases to highlight the difference in magnitude between the results. The columns show the evolution through time at the same periods as Figure \ref{f:stargas} denoted instead by the $t^ \prime$ scale.  }
    \label{f:velocity}
\end{figure*}

As a first order metric of the non-circular velocities in these discs, we show the azimuthal and radial streaming motions of the gas component in a face-on projection in Figure \ref{f:velocity} (also see Appendix\;\ref{a:rc} for the evolution of the rotation curves). The top two rows show the IsoB disc velocity features evolving with time, and the bottom two rows show the TideB disc. A significant difference between the two cases is immediately obvious. It is clear that the tidal forces in the interacting case increase the overall degree of non-axisymmetric motion in the TideB disc. Both velocity components show clear asymmetries that are strongest during closest approach ($t'=-100$\,Myr) which are especially clear in the outer disc whereas the isolated case is relatively unperturbed outside the bar region. These asymmetries correspond with a migration of the system centre of mass, which has been corrected for in Figure\;\ref{f:velocity}. While these strong outer asymmetries in TideB do decay over time as the companion moves away, they remain many times larger than those in the isolated disc. The azimuthal response between the two arms of TideB is also asymmetric, with the northern (southern) arm appearing to rotate somewhat faster (slower) than the disc average, which will result in asysmmetries in galactic shear experienced by gas in each arm region. Only the central regions of the IsoB case seem to show similarly high non-circular velocities which are comparable to the values that fill the entire disc regions of the post-interaction TideB results.

A clear central quadrupole velocity signature is evident in the central region of the radial velocity plots for both discs beyond $t^\prime \ge 0$\,Myr. This feature is indicative of a bar being present in the inner disc, as material streams radially to follow elliptical orbits aligned with the bar (e.g. \citealt{Bovy2019}). The bar kinematics appear strikingly similar between the two models, also showing clear similarities in the azimuthal motion (lower $V_t$ along the bar axis, with two ``lobes" of high $V_t$ perpendicular to the bar axis in the galactic centre, see also \citealt{Renaud2015}). The extent of the quadrupole-like region acts a good visual approximation of the bar extent. It can be seen that the radial size of the bar remains relatively constant once it forms. By qualitatively defining the bar region (as outlined in Section \ref{ss:sf_results}), it was determined that this approximation of the bar extent from the velocity quadrupole was consistent with the values determined from a Fourier method and that the bar extent does indeed remain generally constant once it stabilises over the period of $t^\prime = 200$\,Myr, although the expected bar shortening does become more significant over much longer time-scales. This is also consistent with the assumptions drawn from Figure \ref{f:stargas}, despite the clearly strong differences in the magnitude of non-axisymmetric velocities between the two cases.

One particular point of comparison between the two models is how the velocity features propagate radially. In the IsoB case the positive or negative radial motion traces a single feature from the galactic centre to the outer disc, following from the quadrupole to along the arms and outwards. In the TideB case the feature is not so continuous, with the quadrupole radial velocity features being engulfed by the strong radial motions exhibited by the tidal arms. Such a signature is evidence of the rapid decoupling between the tidal arms and inner bar of the TideB model. Bar and spiral decoupling is a known phenomena in isolated discs (e.g. \citealt{Sellwood1988,Baba2015}) but has not been discussed in the context of tidal interactions where arms are driven by an external agent. We will revisit this topic in Section\;\ref{s:discussion}.

\subsection{Star Formation}
\label{ss:sf_results} 

\begin{figure*}
	\includegraphics[width=\textwidth]{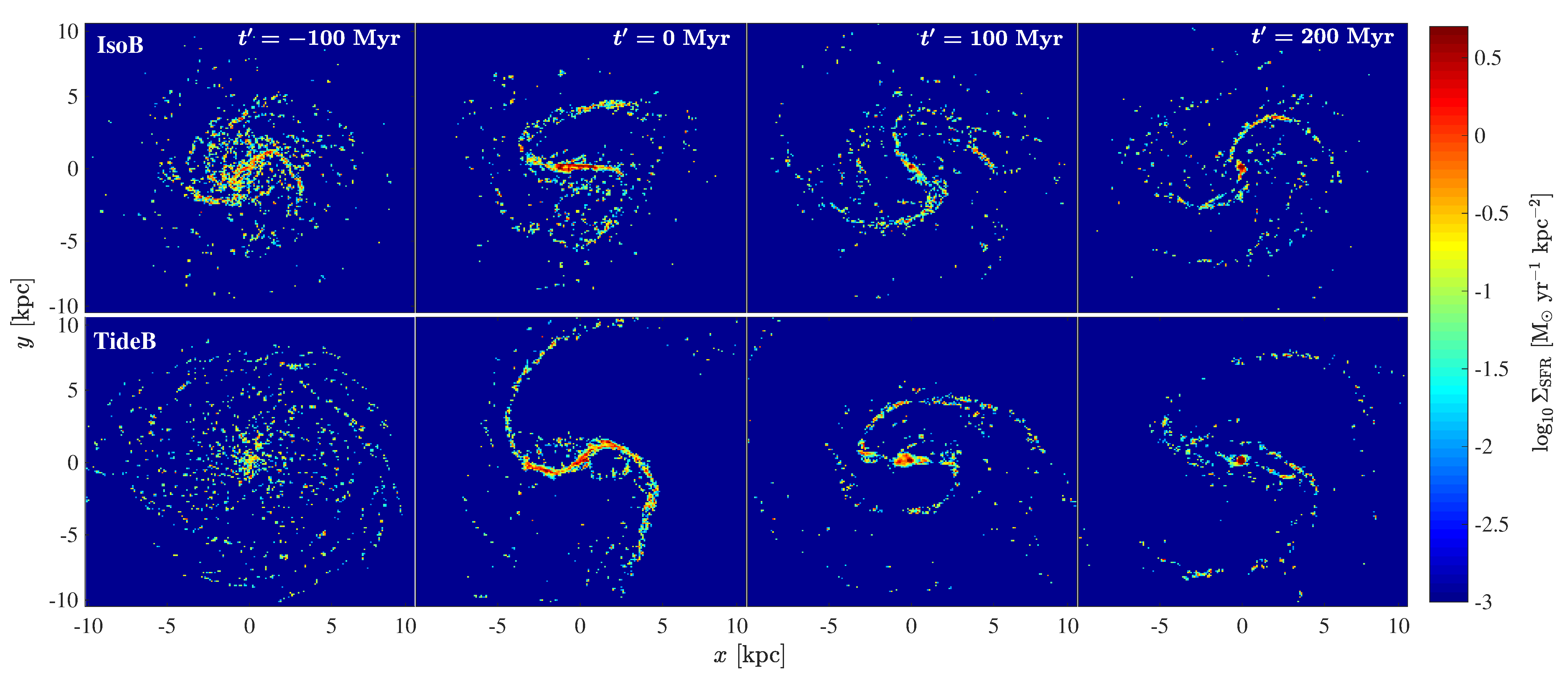}
    \caption{A projection of surface star formation rate ($\Sigma_{\rm SFR}$) into the same $xy$- (disc-aligned) plane for the isolated (top) and tidally driven (bottom) bar models. $\Sigma_{\rm SFR}$ is calculated by counting young stars particles with ages $<\,10$\,Myr and then binning over a 0.1\,$\times$\,0.1\,kpc grid.}
    \label{f:sfr_prj}
\end{figure*}

\subsubsection{Star Forming Morphology} 
We begin by formally constraining the morphology of active star forming regions in the two cases as a measure of the number of new stars formed in a given time period ($\Delta t_{\rm SF}$): 
\begin{equation}
{\rm SFR}(t_i)=\frac{1}{\Delta t_{\rm SF}} \sum_{ t_i}^{t_i - \Delta t_{\rm SF}}  M_{*}
\label{eq:sfr}
\end{equation}
Here, star formation rate (SFR) can be calculated at the current time ($t_i$) by summing new stellar mass ($M_*$) in the star forming period ($\Delta t_{\rm SF}$).This rate is then converted to a surface star formation rate ($\Sigma_{\rm SFR}$) in order to be more comparable to the quantities measured in observations by binning over a 0.1\,$\times$\,0.1\,kpc grid. The look-back time for star formation is the age range $\Delta t_{\rm SF} = 10$\,Myr which is comparable to the star forming time-scale determined observationally for NGC\,4303 \citep{Koda2006,Egusa2009}. A projection of this $\Sigma_{\rm SFR}$ is shown in Figure \ref{f:sfr_prj}. This figure is similarly set into the $xy$- disc plane with columns advancing in time for the same time period as the previous projections of the stellar, gas and velocity features (see Figures \ref{f:stargas}, \ref{f:velocity}). 

From Figure \ref{f:sfr_prj} it is possible to assess the location of current (or very recent) star formation, as well as the intensity of such formation. For example, the centre-most region in all the barred snapshots ($t^\prime \ge 0$\,Myr) for both cases is dominated by an almost circular, bulge-centred region with the highest star formation rate. This circular region expands as the bar evolves in both cases, a likely by-product of gas inflow to this central circumnuclear disc. However, while it is approximately the same size in both cases in the $t^\prime = 100$\,Myr snapshot, by $t^\prime = 200$\,Myr the circular region in the TideB case has become significantly larger than the IsoB case, showing the TideB case has a less concentrated but no less intense region of star formation at the centre. This central nuclear disc of star formation is commonly seen in barred systems, where it is fueled by the inflow of gas from the torques present in the bar region \citep{Athanassoula1992,Wang2012,Cole2014,Baba2020}. Interactions are also known to be drivers of enhanced star formation (e.g. \citealt{Patton2013,Pan2019}), with the interaction event itself driving inflows that help fuel central star formation \citep{Torrey2012,Pettitt2016}. As such, the combined effect of the bar evolution and the interaction provides an increase in fueling of the central nuclear disc in TideB, and the larger central star forming region.

The bar is also a clear feature in these SFE maps, with pockets of intense star formation flaring along the bar and in clumps along the spiral arms. However, star formation for the tidally driven bar appears to be predominantly contained within the bar and arm features in all barred time periods ($t^\prime > 0$). In contrast, each of the IsoB maps in Figure \ref{f:sfr_prj} show wide areas of strong star formation throughout the disc which cannot be clearly associated with either the bar or any obvious arm feature. This seems to indicate the isolated IsoB case has a greater amount of inter-arm star formation, whereas the tidally driven case concentrates gas more strongly into the arms and core, limiting star formation outside of these areas. 

Considering time dependent features, the shape of the SFR projection in the bar region changes in both cases dramatically over a relatively short time-scale ($\Delta t^\prime \approx 200$\,Myr). Figure \ref{f:sfr_prj} shows distinctly different star forming structures within the bar region, and that the evolution process of this structure can be seen to differ between the two cases. For example, in the $t^\prime = 200 $\,Myr snapshot of the IsoB case, very clear arms comprised of star forming regions can be seen to be somewhat disconnected from a dense, almost triangular central region within the bar. Comparatively, in the same snapshot for the TideB case, the central region is also clumpy but is mostly connected to the arms along a straight central axis of star formation that spans the full bar length. This axis is accompanied by a curved envelope of similarly intense star formation encasing the central axis from both above and below, which seemingly connects each arm to the other. Additionally, the bar in the TideB model at $t^\prime = 100 $\,Myr shows a wide, long clump along the bar axis in the centre which tapers off toward the middle of the bar, before once again intensifying into a thicker bar at the end just before the arms connect. This is dissimilar to any of the other snapshots for either case. These star forming structures vary broadly, even between evolutionary periods in a single simulated case which make it difficult to produce any sweeping, general statements to describe definitive features. However, this difficulty itself supports the range of observations that indicate star formation within galaxies seems to be consistently inconsistent.  

\subsubsection{Star Forming History}
\label{sss:sf_history}
\begin{figure}
	\includegraphics[width=\columnwidth]{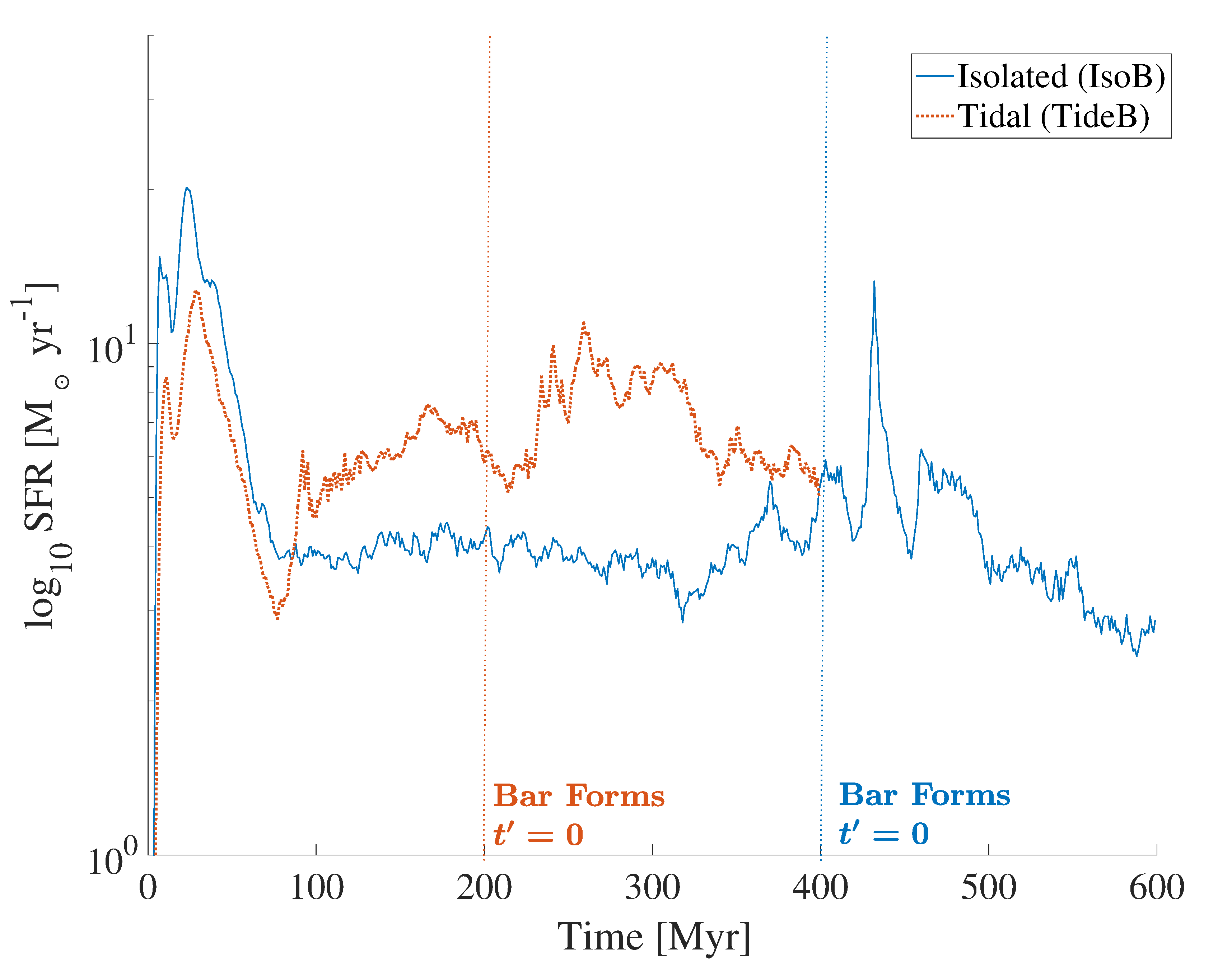}
    \caption{Star formation history for the two cases for the evolutionary period until $t^\prime = 200$\,Myr. The axis is set at original simulation time so, for each case, a vertical line denotes $t^\prime = 0$\,Myr. The solid blue line represents IsoB. The orange dotted line represents TideB. The large peak at $t<100$\,Myr is an artefact of the simulation process. Vertical dotted lines approximately indicate bar formation times.}
    \label{f:sf_history}
\end{figure}

To consider more specifically the time evolution of star formation in the simulations, we take the disc-averaged star formation history for the two simulations up to $t^\prime = 200$\,Myr and present Figure \ref{f:sf_history}. The axis in this figure is set in the original simulation time, and for each case a vertical line denotes $t^\prime = 0$. The solid blue line is the star formation history for the isolated IsoB case, whereas the orange dotted line indicates the history of the tidally driven case. The time of closest approach for the companion in this case is at approximately 94\,Myr. 

It is apparent that both cases seem to show relatively similar star formation rates. There are also some similarities in the shape of the star formation profile as it evolves. For example, there is evidence of an extended rise in star formation at around the time the bar forms in both cases (indicated by the vertical dotted lines). The relative shape (height and duration) of this increase in both cases appears to be similar: IsoB $ \sim \Delta \mbox{\rm SFR} = 3.08$\,M$_\odot$\,yr$^{-1}$, $\Delta t = 145$\,Myr; TideB $\sim \Delta \mbox{\rm SFR} = 3.79$\,M$_\odot$\,yr$^{-1}$, $\Delta t = 127$\,Myr. The first obvious difference between the two cases is in the earlier stages of evolution (pre-bar formation $t^\prime \le 0$\,Myr). While the IsoB case seems to have a relatively constant level of star formation before the bar forms (after the initial boost on start-up), the TideB case has an additional, earlier period with an extended rise and fall of star formation intensity which is similar to the boost feature previously linked to the bar formation. This occurs just after the closest approach of the companion. Although this early boost is smaller in amplitude than the bar effect, it appears to last for a similar duration before returning to what seems to be the original baseline rate just before the advent of the bar formation, where it steeply rises again. This can be considered to be a double peaked trend of star formation history and is likely attributable to a starburst triggered by the interaction. As the bar has not yet formed and the density and SFR projections both show the two arm features are dominant in that period, a logical assumption would be for this boost to be mostly limited to increased star formation in the strong tidally induced arms, rather than within the central region which later becomes the bar. The magnitude of the bar-triggered boost in SFR is similar for both the isolated and perturbed disc (79\% and 66\% respectively), while the interaction triggered burst is milder at 31\%.

To study the morphological dependence of these star formation features in more depth, three key morphological regions are defined to be the bar, arm and inter-arm components. Using Fourier decomposition, it was possible to identify analytically relevant traits of the bar and arm components. The inter-arm region could then be defined as neither bar nor arm within a certain radius of interest. The extent of the bar region was determined from the amplitude of the Fourier $|A_2|$ mode. Identifying the radial extent of features in this mode allowed for the bar component to be defined with a simple radial criterion to form a central circular region encasing the full bar length.

\begin{figure}
	\includegraphics[width=\columnwidth]{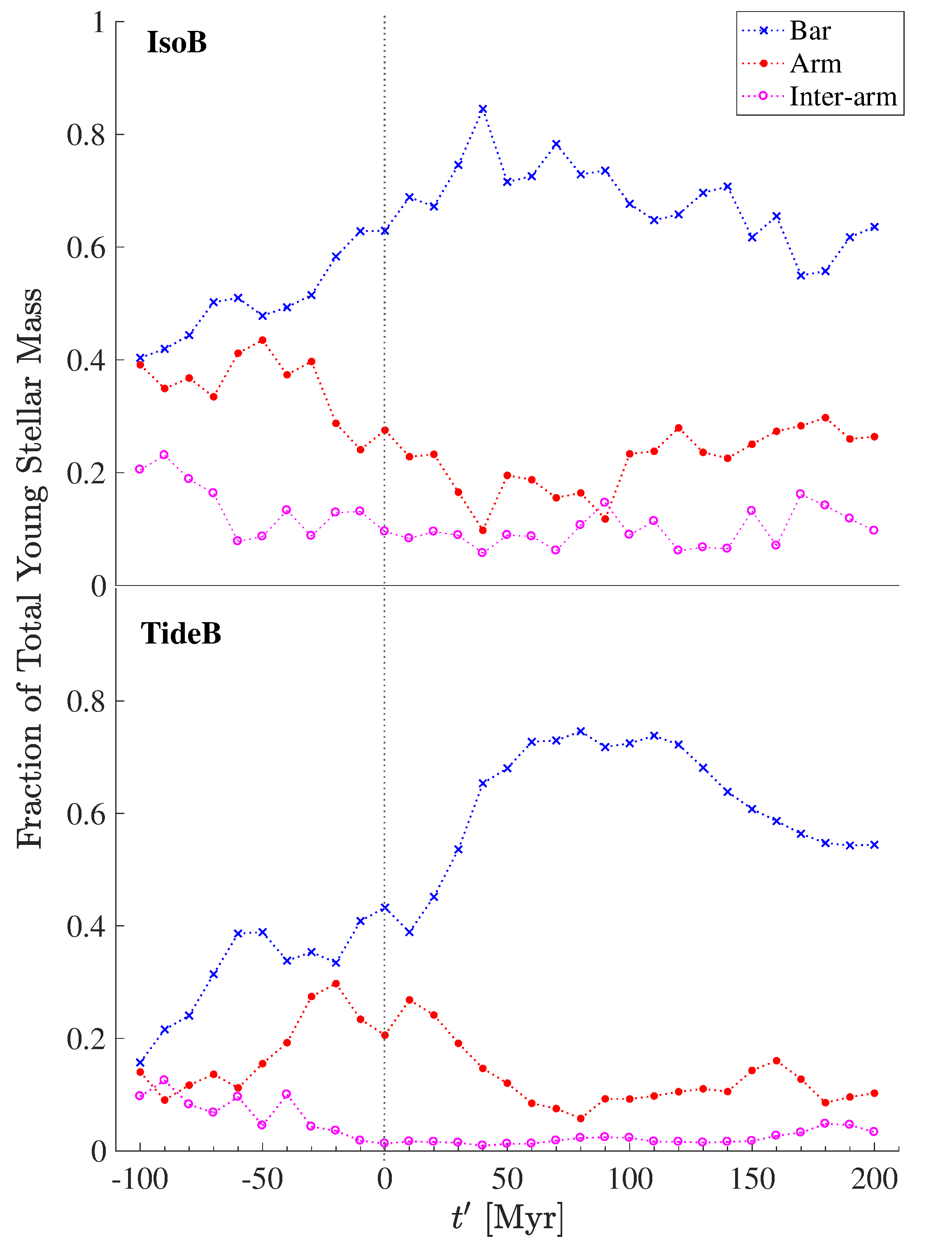}
    \caption{The fraction of total young stellar mass produced by each region (bar, arm, inter-arm) over the period of interest ($-100\,<\,t^\prime\,<\,200$\,Myr) for each bar model. Each region is only defined on steps of 10\,Myr, corresponding to the previously assumed star formation time. The bar is denoted by a blue line with crosses at each calculation point; the arms by a red line with filled-in circles; and the inter-arm with magenta line and open circles. The axis is set in the $t^\prime$ scheme with a dotted vertical line at $t^\prime = 0$ to indicate the approximate time of bar formation.  }
    \label{f:sf_regions}
\end{figure}

As mentioned previously, it was found that the bar extent was relatively constant throughout the considered duration, decreasing only a small fraction by $t^\prime = 200$\,Myr. Hence, a single value was adopted for the bar extent for the following analysis (IsoB: $R_{\rm bar}=2.10$\,kpc; TideB: $R_{\rm bar}=3.12$\,kpc). From the population outside of this $R_{\rm bar}$, the arm component was defined by fitting a log spiral of the form $r=a\exp{b\theta}$ with a standard width of $\pm 1$\,kpc to the peaks of the polar angle $\theta$ set to lie in the disc plane at a given radius. The inter-arm region was correspondingly defined as the remainder within a set radius of 10\,kpc determined to contain the majority of significant disc features for both cases. 
 
In Figure \ref{f:sf_regions}, a similar star formation history is plotted to Figure \ref{f:sf_history}, however, this is a measure of the fractional contribution from each region (bar, arm and inter-arm) to the total number of young stars produced, shown over the simulation time. These specific regions were defined and analysed at every 10\,Myr, with a look-back time consistent with the previous star formation analysis ($\Delta t_{\rm SF} = 10$\,Myr). The blue line with crosses at each calculation point denotes the contribution from the defined bar region; the filled-in red circles indicate the contribution from within the arm regions; and finally, the open circles in magenta show the inter-arm contribution. The axis is set in the $t^\prime$ scheme with a vertical line at $t^\prime = 0$ to indicate the approximate time of bar formation.  

\begin{figure}
	\includegraphics[width=\columnwidth]{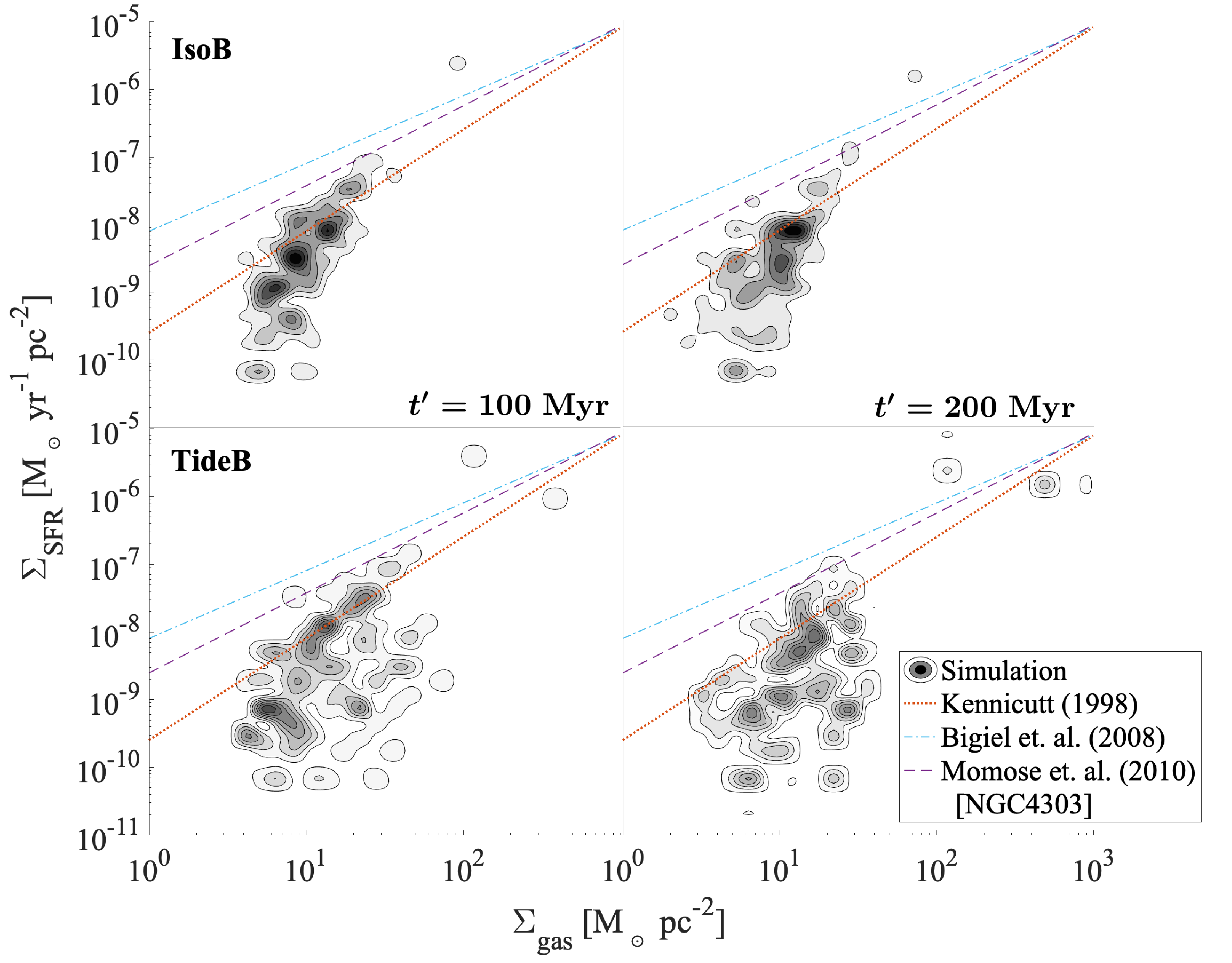}
    \caption{The Kennicutt-Schmidt relation between surface star formation rate ($\Sigma_{\rm SFR}$) and surface gas mass density ($\Sigma_{\rm gas}$ ) for both cases during the clearly barred periods $t^\prime = 100, 200$\,Myr. These surface densities are determined as in Figure \ref{f:sfr_prj} and smoothed with a spline interpolation. The literature relations: \citet{Kennicutt1998} orange dotted line; \citet{B2008} cyan dot-dashed line; and \citet{Momose2010} purple dashed line--are each shown for comparison.}
    \label{f:ks_literature}
\end{figure}

In both cases, it is possible to see how star formation in the bar region climbs steadily as the bar is formed and that the inter-arm region contributes the least for the majority of times analysed. For the IsoB case, this inter-arm contribution is consistently $\sim10\%$ of the total stellar mass formed and the fraction is even lower in the TideB case ($\sim 5\%$ or less once the bar has formed). This is also a representation of the way that the star formation seems to be more cleanly located in the apparent bar and arm features in the TideB case, particularly as seen in Figure \ref{f:sfr_prj}. Considering the contributions from both bar and arm components however, there are a number of very noticeable differences between the two cases. For example, for times when $t^\prime \le 0$\,Myr, it can be seen that the arms in the TideB case show a very different development of the fraction of stars produced. In the IsoB case, the fraction of star formation over the disc features seems to be initially similar for the first $\sim50$\,Myr ($t^{\prime} \le -50$\,Myr), before becoming increasingly centrally dominated as the bar forms ($\sim t^{\prime} = 0$\,Myr). Contrastingly, the highest period of intensity for star formation in the TideB arms is very noticeably within $\pm\,50$\,Myr either side of the bar formation time ($-50 \le t^{\prime} \le 50$\,Myr). This corresponds to the time in Figure \ref{f:sf_history} where the smaller peak of star formation can be seen after closest approach and before the bar has completely formed. Compared to the IsoB case, which has a steady baseline level of star formation at that time, the interaction has clearly driven a burst of star formation which is predominantly located in the arms. At later times in Figure \ref{f:sf_regions} ($\sim t^\prime > 50$\,Myr), a potential periodicity also appears in the contribution from the bar region of the IsoB case which is not reflected in the bar of the TideB case. The TideB case at this time shows a much smoother profile. While overall star formation history may appear relatively indistinguishable once the bar has formed, a number of significant differences seem to exist when considering instead the contribution to the overall SFR from region to region.

\begin{figure}
	\includegraphics[width=\columnwidth]{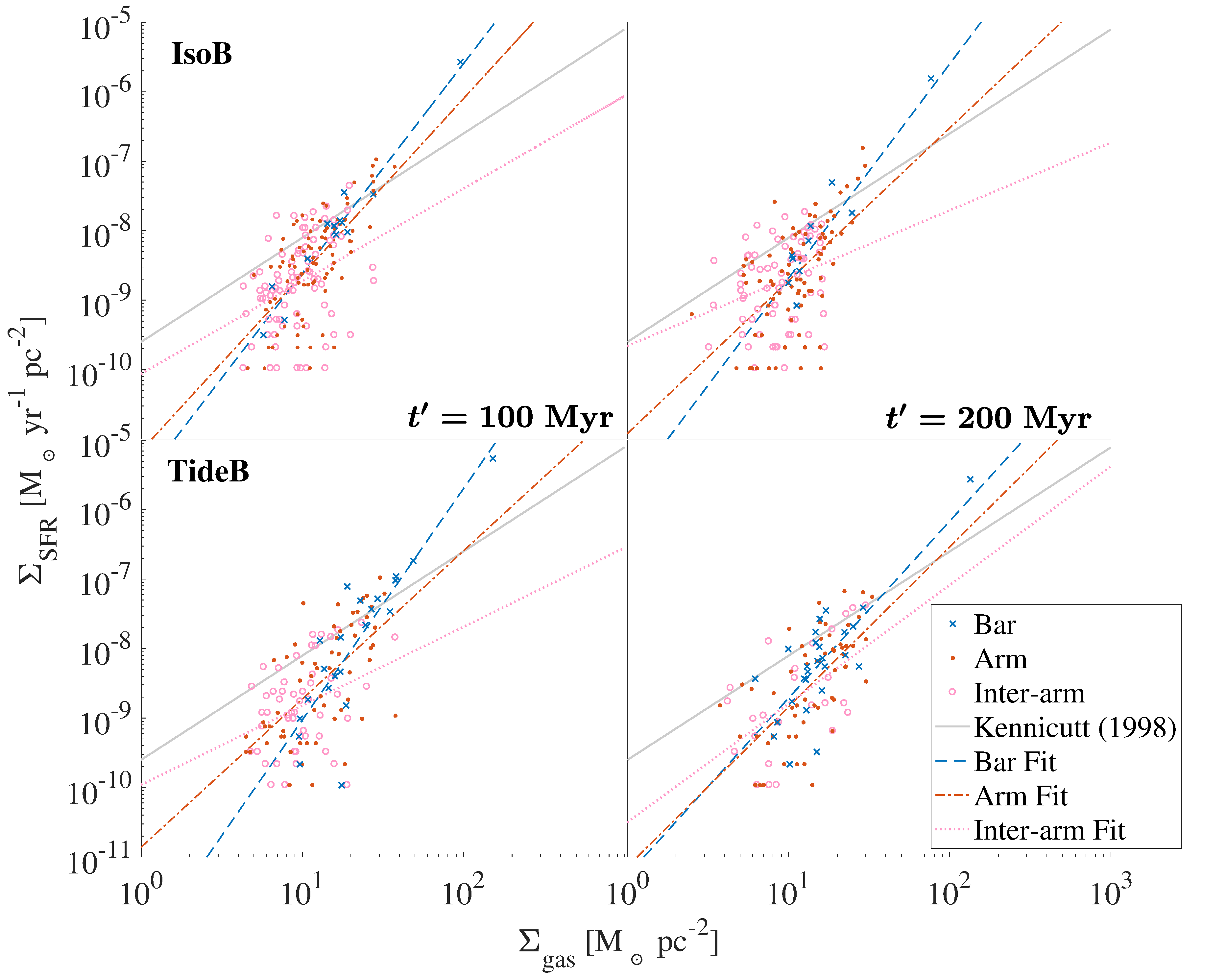}
    \caption{As in Figure \ref{f:ks_literature} but now decomposed by region with scattered data points showing the bar (blue crosses), arm (red filled circles) and inter-arm (magenta open circles) components. The \citet{Kennicutt1998} relation is again included for reference (dotted orange line), as well as power law fits for each region: bar (blue solid line), arm (red dashed line) and inter-arm (magenta dot-dashed line).}
    \label{f:ks_regions}
\end{figure}

\subsection{Star Formation Scaling Relations}
To compare our measured star formation analytically with observations, as well to consider these results in relation to the gas availability rather than simply counting stars, we plot a range of the so-called Kennicutt-Schmidt diagrams showing the relation between $\Sigma_{\rm SFR}$ and surface gas mass density ($\Sigma_{\rm gas}$) \citep{Schmidt1959,Kennicutt1998}. These can be seen in Figure \ref{f:ks_literature} and Figure \ref{f:ks_regions} matching the resolution used in Figure \ref{f:sfr_prj} (binning over a 0.1\,$\times$\,0.1\,kpc grid). For both simulated cases, Figure \ref{f:ks_literature} shows this relation in a plot for each of the two clearly formed bar snapshots ($t^\prime = 100, 200$\,Myr). These are compared with literature relations derived from observational studies. The calculated data points in this $\Sigma_{\rm SFR}$ vs. $\Sigma_{\rm gas}$ are shown as contours, whereas the coloured lines are the literature relations: \citet{Kennicutt1998} in orange dotted; \citet{B2008} in cyan dot-dashed and \citet{Momose2010} in purple dashed. \citet{Kennicutt1998} use disc averaged values for many galaxies; \citet{Momose2010} use only values from across NGC\,4303; and \citet{B2008} constrain their measurements to be \textquoteleft resolved\textquoteright \ star formation in many galaxies. 

The shape of the distribution is clearly different between the two cases. The IsoB case has a narrow, arrowhead-like shape which seems to align relatively well in orientation with the literature results. Comparatively, the TideB case is much broader along the $\Sigma_{\rm gas}$ axis making more circular distribution with less clear alignment. In TideB in general there are many more components with high $\Sigma_{\rm gas}$ and low $\Sigma_{\rm SFR}$. This indicates TideB appears to have noticeably more regions over the disc which are relatively inefficient at forming stars, but can maintain higher gas density. Both cases show there is clear evidence of evolution with time, in terms of both shape and location of the distribution of simulated results with respect to the literature results. This is particularly evident in the area of the plot which corresponds to low efficiency, high gas density features (around $\Sigma_g>10\,{\rm M_\odot pc^{-2}}$ and $\Sigma_{\rm SFR}<10^{-9}\,{\rm M_\odot pc^{-2}yr^{-1}}$) which seem to become more pronounced, spreading to higher gas densities for a given $\Sigma_{\rm SFR}$ value, as both cases evolve. However, while the TideB distribution in Figure \ref{f:ks_literature} shows many more such  high $\Sigma_{\rm gas}$ and low $\Sigma_{\rm SFR}$ results than IsoB, the significant difference between the two cases in this region is conspicuously absent when these results are are deconstructed into components of the three morphological regions of bar, arm and inter-arm (see Figure \ref{f:ks_regions}). This indicates that many of these points in the TideB case were not classified as either bar, arm or inter-arm through the classification scheme, indicating the presence of some other component. 

While these values trace star formation within the $10 \times 10$\,kpc field of view that can be seen in the windows of the face-on galaxy maps, these points must be outside the $R\,>\,10$\,kpc radial limit used in the region classification to define the extent of the optical disc. These results should then be attributed to star formation occurring in high density gas which has been stripped to the very outer edges of the disc or even into a trailing tidal tail by the companion. This shows that the largest temporal changes in the TideB case are likely to be a direct result of the interaction still affecting the outer-reaches of the optical disc and beyond. Additionally, that the definition of the disc limit may make significant impact on the shape of the KS relation in tidally affected galaxies. In response, slightly changing the outer radius of the disc for both cases was also tested to find that the overall shape of the IsoB distribution is mostly unaffected while the TideB distribution is clearly sensitive to the change. As we are predominantly concerned with the visible morphological features within the disc (bar, arms, inter-arm) this is not considered in great detail, however, studies of stellar populations in the outer discs of galaxies or tidal features in interacting systems may find such sensitivity particularly relevant. 

In comparison, Figure \ref{f:ks_regions} is similar to Figure \ref{f:ks_literature} but the simulation points are deconstructed into the components of bar, arm and inter-arm, as defined previously. For the purpose of producing this relation, the definition of morphological features is based on the $(x,y)$ positions determined previously such that these regions in both gas and stellar components are spatially consistent. These are presented through scattered points, identifiable with the same symbols as in Figure \ref{f:sf_regions} (bar with blue crosses, arms with filled red circles, inter-arm with magenta open circles). The \citet{Kennicutt1998} relation is included again for reference as in Figure \ref{f:ks_literature}, now with a grey solid line. This allows us to begin to understand how each region contributes to the overall distribution, as well as to make an assessment of how the star formation efficiency may differ from region to region. For example, in observations of NGC\,4303, \citet{Momose2010} found the overall star formation efficiency to be higher than the average trend for all galaxies derived by \citet{Kennicutt1998}, particularly in the arm regions. That is not necessarily the case here, although a  fraction of the arm results do show values higher than the \citet{Kennicutt1998} relation.

The most interesting result is that in this figure the bar in both cases has the steepest slope, followed  consistently by the arm and then inter-arm regions. While visually this is clearer for some panels than others, with the early IsoB data appearing to show similar slopes for the arm and bar, through  calculations of the gradients these trends are obvious. Table \ref{t:KS_fits} shows the $\alpha$ coefficient for a power law fit to the simulation data of the form $\log \Sigma_s = \alpha\log \Sigma_g + \beta$ for each region component as well as the total distribution over the disc ($R<10$\,kpc as in Figure \ref{f:ks_regions}). Considering these coefficients as an analogue for the linear gradient in the log-log space of Figure \ref{f:ks_regions}, it can be confirmed that the bar region in both cases certainly shows a significantly steeper relation than the arms, inter-arm or even total distribution on average. 

\begin{table}
	\centering
	\caption{Gradient coefficients for a power law fit to the previous $\Sigma_{\rm gas}$ vs. $\Sigma_{\rm SFR}$ data for the disc region of interest ($R<10$\,kpc). This fit follows the form $\log \Sigma_s = \alpha\log \Sigma_g + \beta$, the $\alpha$ coefficient is listed.}
	\label{t:KS_fits}
	\begin{tabular}{lcccc} 
		\hline
		\bf{Model -- $t^{\prime}$} &  \bf{All} & \bf{Bar} & \bf{Arm} & \bf{Inter-arm} \\
		 \hline
		IsoB-100\,Myr & $2.7\pm0.4$ &  $3.0\pm0.5$ &  $2.5\pm0.6$ & $1.3\pm0.8$\\
		IsoB-200\,Myr  & $2.3\pm0.4$ &  $3.1\pm0.7$ &  $2.2\pm0.7$ &  $1.0\pm0.8$  \\
		TideB-100\,Myr & $2.5\pm0.4$ &  $3.0\pm0.9$ &  $2.4\pm0.7$ &  $1.6\pm0.9$ \\
		TideB-200\,Myr  & $2.4\pm0.5$ &  $3.2\pm0.8$ &  $2.4\pm0.7$ &  $1.7\pm1.1$ \\
		\hline
	\end{tabular}
\end{table}

The outlying point with highest $\Sigma_{\rm SFR}$ and $\Sigma_{\rm gas}$ can be seen to always correspond to the bar component. In fact, this is the result from the centre-most region which is defined in some studies as a fourth morphological feature (a central nucleus) and is considered separate from bar properties. For the subsequent analysis, care has been taken to test both including and excluding this point in assessing the bar response. The results included in Table \ref{t:KS_fits} are for fits including the nucleus within the barred region and excluding star formation at and beyond the defined disc limit of 10\,kpc. However, the addition of the nucleus has been determined to make negligible change with the bar slope without this extremely high value still significantly steeper than all other results. The addition of tidal remnants in the total average, however, saturates the overall fit to be more consistent with the flatter inter-arm component rather than the arms. This is expected as within the disc star formation in the arms spatially dominates the total star formation, particularly in the TideB case as can be seen in Figure \ref{f:sfr_prj}, whereas the conditions in the outer regions are more consistent with the inter-arm space. Hence, when these outer regions are included the influence of the arms on the total average must be decreased. 

Comparing these results with related studies, it is possible to see that the range of results is relatively similar. In the case of the isolated result, \citet{Momose2010} also find a similar arrowhead shape for NGC4303, however it is possible to see from Figure \ref{f:ks_literature} that the alignment of the distribution is steeper overall than the relation observed by \citet{Momose2010} with the IsoB result showing consistently lower values for star formation in the low gas region of the plot, causing a steeper alignment of the arrow shape overall. Additionally, when considering the regional separation, observational results appear to indicate a preference for a significant portion of the arm component to have distinguishably higher SFE than the other regions; with arm values independently occupying the top arrow-edge and showing a wider spread of values to higher SFR for a given gas value than any other regional components. The bar population is similarly shown to have lower SFR than either the arm or inter-arm component for higher gas surface densities. Such separable features are not necessarily evident in the IsoB result, firstly as arm and inter-arm are indistinguishable, and secondly because while the arm and inter-arm regions show higher SFR than the bar for similar gas density they also show lower values at the same gas value. In the IsoB results, the bar component forms a thin distribution almost bisecting the larger spread of the other regions. Comparatively, in the TideB case, the bar response is more dispersed throughout the distribution, particularly in the second time period ($t^\prime=200$\,Myr). This is in line with the results of \citet{Watanabe} where the region of the KS plot occupied by bar results is also predominantly interspersed with the other region. 

However, it is important to note that \citet{Watanabe} do not specifically define an arm and inter-arm region in their sample, leaving these populations somewhat ambiguously in a region defined as \textquoteleft Other (mostly arms)\textquoteright. There is also a difference in the bar region definition as \citet{Watanabe} define both a bar region plus a separate classification for the bar ends which neither \citet{Momose2010} or this work have elected to differentiate from the bar on the whole. It is very likely that the difference between region classification has introduced a certain level of ambiguity into the possible comparison between regionally specific trends between studies. In addition, \citet{Onodera2010} have shown that resolution in observational studies can have a non-trivial affect on the shape and orientation of the KS relation even within a single target galaxy. By nature, simulation studies can achieve a significantly higher resolution than most observational results. Even within simulations, for example, \citet{Fujimoto2014} have shown that the KS relation is sensitive to the star formation models used to prescribe the physics of the simulation. The differences are particularly evident when considering the regionally dependent relation with standard star formation as prescribed by mass and free-fall time constraints, GMC turbulence as per \citet{Krumholz2005} and cloud-cloud collision as per \citet{Tan2000} each producing a different orientation of the same bar, arm and inter-arm regions in $\Sigma_{\rm gas}$--$\Sigma_{\rm SFR}$ parameter space. The results presented here are likely not immune to such constraints. For example, the IsoB results in Figure \ref{f:ks_regions} are similar to the standard response in \citet{Fujimoto2014}--unsurprising considering the star formation recipe used in this simulation--whereas, the observational results of \citet{Momose2010} appear to be more similar to the cloud-cloud collision result of \citet{Fujimoto2014}. Hence, it is possible that some discrepancy between these results and the observational results from both target galaxies may arise due to inherent resolution and star formation conditions, as well as the region classification method which is consistently inconsistent in the literature.   

Considering instead the mean values for $\Sigma_{\rm{SFR}}$ and $\Sigma_{\rm{gas}}$ for each region, the corresponding values show an evident decrease from bar to arm to inter-arm which is interestingly consistent with the distinct flattening of the slope for the $\Sigma_{\rm{gas}}$--$\Sigma_{\rm{SFR}}$ relation (as in Figure \ref{f:ks_regions}, Table \ref{t:KS_fits}). This is true in all time periods. The gas means also reflect decreasing values per region in order of bar -- arm -- inter-arm but the magnitude of the difference between each region is less in the gas surface density than the star formation. These mean values for each of the defined regions are shown in Table \ref{t:KS_means} for each model. However, in this classification the high $\Sigma_{\rm{SFR}}$, high $\Sigma_{\rm{gas}}$ outlying point attributed to the central nucleus is included within the bar component. If this is removed and the averages for the originally defined regions are calculated instead as four regions, it can be seen that this nucleus saturates the response of the bar. Table \ref{t:KS_means} shows the original bar component with nucleus as Bar and the bar without the centre-most region as Bar (/cen.). The Centre result is exactly the points removed from the original bar component. Of the four regions, the nucleus is considerably the highest by two orders of magnitude, however, the bar without centre still produces a value significantly higher than the corresponding arm region which is in turn greater than the mean inter-arm values. This is also consistent with the related analysis of simulated discs by \citet{Fujimoto2014} who similarly find  in the disc decomposed by region that the average of bar values are generally higher than the average arm region and then the inter-arm region with varying degrees of separation depending on the star formation model implemented.  

\begin{table}
	\centering
	\caption{Mean values of gas and star formation surface density in morphologically distinct regions in each model.}
	\label{t:KS_means}
	\begin{tabular}{lcccccc} 
		\hline
		{\bf{Region}}&  \multicolumn{2}{c}{\bf{$\bar{\Sigma}_{\rm gas} \,{[\rm M_\odot pc^{-2}]}$}} &  \multicolumn{2}{c}{\bf{$\bar{\Sigma}_{\rm SFR} \,{[10^{-2} \rm M_\odot yr^{-1}pc^{-2}]}$}} \\
		& \bf{IsoB} & \bf{TideB} 		& \bf{IsoB} & \bf{TideB} &\\
		 \hline
        Arm     & $12.9\pm0.6$ & $15.3\pm0.2$  & $1.01\pm0.1$ & $1.14\pm0.3$\\
        Bar     & $20.1\pm1.1$ &  $23.6\pm5.4$ & $16.2\pm5.8$ & $19.6\pm12$\\
        Centre     & $86.1\pm14$ & $143\pm13$  & $212\pm79$ & $410\pm193$\\
        Bar (/cen.) &$15.0\pm0.1$ & $18.4\pm4.3$ & $1.13\pm0.1$ & $2.17\pm1.7$ \\
        Inter-arm  & $10.0\pm0.5$ & $12.0\pm1.4$ & $0.42\pm0.1$ & $0.56\pm0.3$ \\
		\hline
	\end{tabular}
\end{table}

Comparing these results to observationally determined values for the average SFR surface density, as calculated by \citet{Yajima2019} for NGC4303 and \citet{Watanabe} for NGC3627, indicates that $\Sigma_{\rm{SFR}}$ for the main features of bar (not including the nucleus) and arms are at least of a similar order. The simulation values are all slightly lower than the observational counterparts for each region. Comparing the relative values between the two main morphological regions, \citet{Yajima2019} find that the arms in NGC4303 have a higher $\Sigma_{\rm{SFR}}$ than the bar by $\sim$2\%. For NGC3627, \citet{Watanabe} indicate that the bar should have $\sim$4\% higher $\Sigma_{\rm{SFR}}$ than the other region (classified as \textquoteleft mostly arms' so, possibly also including an inter-arm component). This is a very small difference between the two regions in observations of both targets. However, both these studies define and calculated $\Sigma_{\rm{SFR}}$ while separating the bar ends to a 3rd significant region which has significantly higher value than either the arms or bar. In this study, the bar ends are not specifically separated and it is difficult to know to which component (bar or arms) in our classification that this 3rd region has been drawn from to compare with the observational studies. 

Assuming the bar and bar ends together make a complete bar component, averaging the bar and bar end results from observations \citep{Watanabe,Yajima2019} instead indicates the bar+bar end component of NGC4303 to have a $\Sigma_{\rm{SFR}}$ $\sim27$\% more than the arms whereas this is $\sim$71\% for NGC3627 with the bar+bar end component being higher than the arms in both cases. This resolves the difference between the IsoB simulation showing lower star formation in the bar rather than the higher star formation rates in the arms than the bar of NGC4303 \citep{Momose2010,Yajima2019}. The simulation results also show a significantly greater difference between the bar and arm components in the TideB result than IsoB, similar to the corresponding targets, although the calculated values are also lower than observations in this respect. The IsoB case has the bar (/cen.) greater the arm by 11\% whereas the TideB case has a difference of 47\% between the two regions. However, as mentioned previously, the spatial definition of regions is often inconsistent between studies (both observational and theoretical) and this inconsistency can give rise to significant differences in the trends observed. So, in the following section we attempt to quantify how the star formation properties differ spatially throughout the bar and disc in a method independent of a specific regional classifications. 

\subsection{A Directional Approach}
\label{ss:detailed_results} 
To further investigate the sub-kpc star formation features of these galactic bars, we perform an analysis of the SFE within the inner regions of each disc. We apply a directional approach to consider how star formation varies and evolves with disc radius, as well as along the primary axis of the bar--a long contested topic in the literature (e.g. \citealt{Roberts1979,Muraoka2019}). Star formation efficiencies are used as a metric specifically to account for the interrelation between gas availability and star formation activity, with SFE $= \Sigma_{\rm SFR}/\Sigma_{\rm gas}$ [Myr$^{-1}$].

\subsubsection{Radial Dependence of SFE}
\begin{figure}
	\includegraphics[width=\columnwidth]{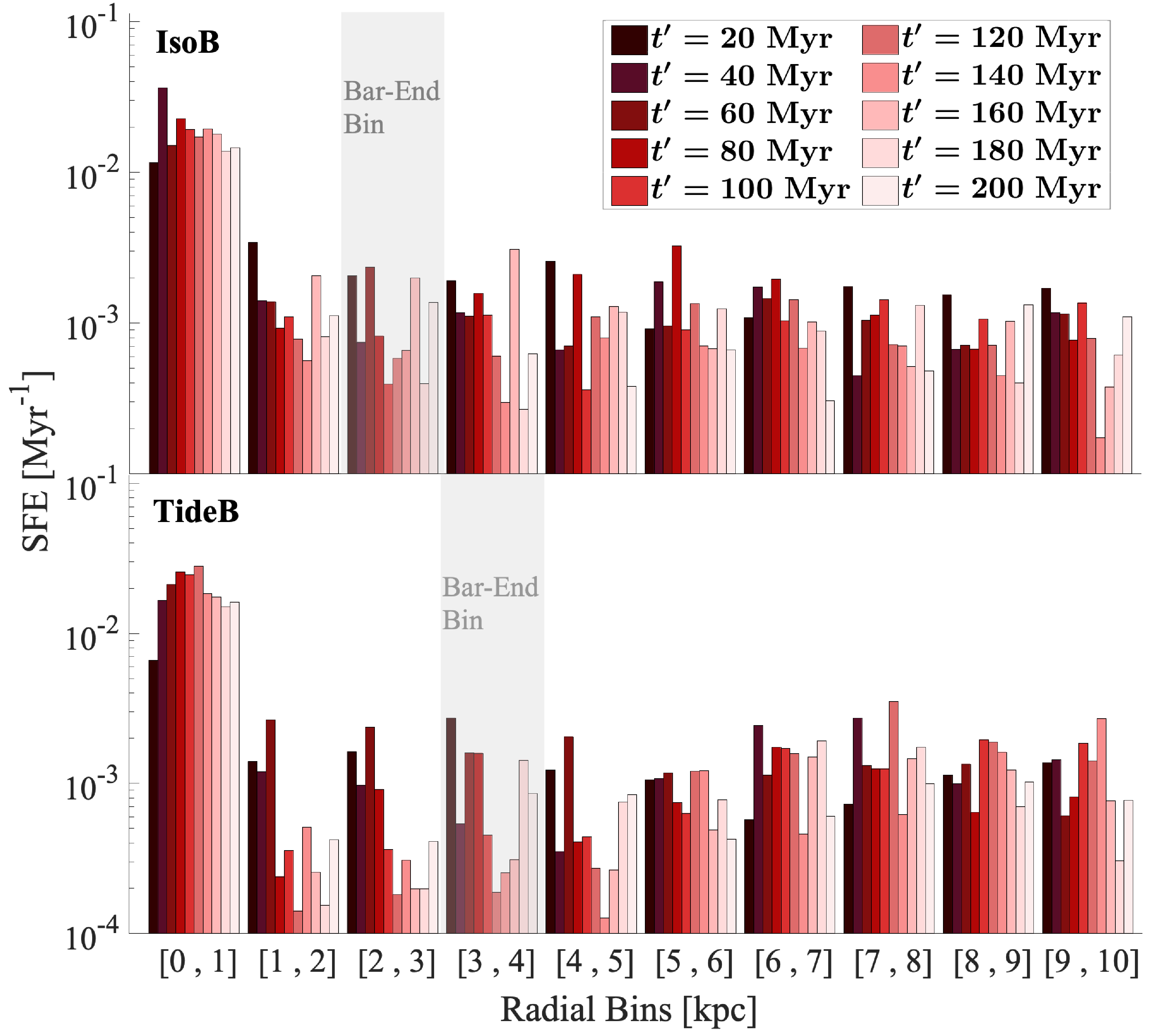}
    \caption{Azimuthally averaged SFE with radius for each bar model (IsoB top, TideB bottom). The SFE across the disc is binned in 1\,kpc steps of radius for the barred period ($t^\prime > 0$). The star formation time remains consistent with previous analysis ($\Delta t_{\rm SF} = 10$\,Myr). For each radial bin, the evolution of SFE can be seen with individually coloured bars corresponding to the result at time steps of 20\,Myr. The colour gradient shows time evolution, with dark to light bins corresponding to forward time progression (over 200\,Myr). The bin in which the bar extent would be located is shaded in grey.  }
    \label{f:hist3_Rsfe}
\end{figure}

The radial dependence of SFE across the disc of each case is shown in Figure \ref{f:hist3_Rsfe}. Each radial bin is defined with a width of $\Delta R = 1$\,kpc and the SFR and gas mass averaged over each annuli to calculate the SFE value in each bin. This is determined every 10\,Myr for each case over the barred period ($t^\prime > 0$\,Myr) with consistent star formation look-back time ($\Delta t_{\rm SF} = 10$\,Myr). Saturation indicates time evolution with colour from dark to light corresponding to increasing $t^\prime$. Additionally, the bin which would coincide with the bar end is shaded in grey.  

Naturally, the centre-most bin shows the highest result at all times for both cases of bar formation. This is in line with the noticeably high star forming central region, or nucleus, which was identified in previous sections (see Figure \ref{f:sfr_prj}, \ref{f:ks_literature}, \ref{f:ks_regions}). Within each bin, there does not appear to be any immediately obvious or consistent trend as the SFE varies with time in either case. It neither continuously grows, declines, nor oscillates in any discernible pattern. In considering the broader trend as SFE changes with radius across all bins, there does appear to be a certain difference between the two cases. Outside of the central bin, the average level of SFE in the disc appears to be mostly constant in the IsoB case, whereas the TideB case shows more of an increasing or sinusoidal shape with radius; showing lower SFE towards the centre ($\sim R_{\rm bar} \pm 2$\,kpc) and higher SFE towards the outer edge. This may be evidence that the interaction has caused the inner regions to suffer a depletion of gas compared to the relatively more stable conditions in outer disc. However, these broad trends are not necessarily true at all times, as the inner deficit is clearly not prevalent in the first 3-4 periods ($\sim t^\prime < 80$\,Myr).

\begin{figure}
	\includegraphics[width=\columnwidth]{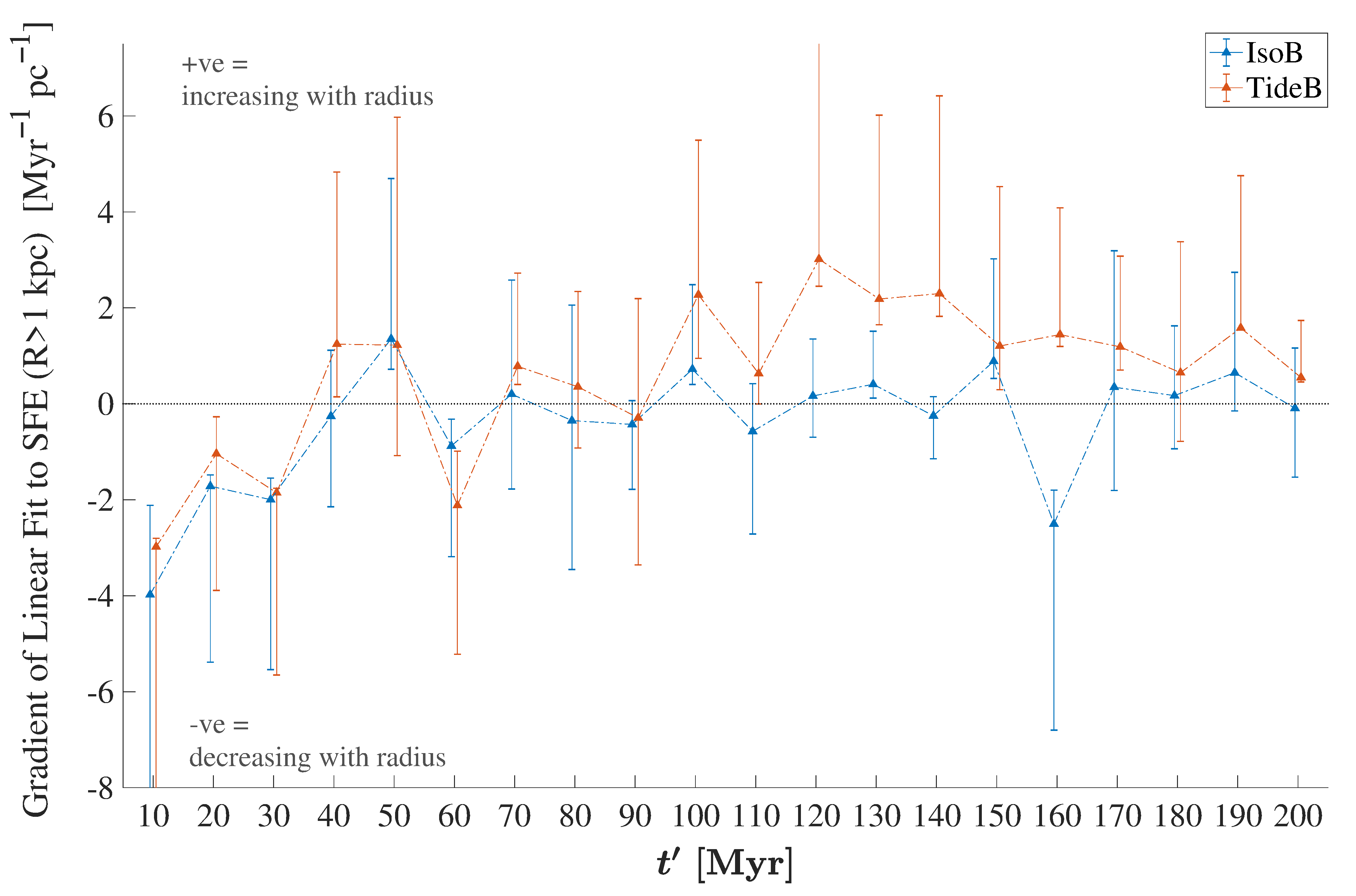}
    \caption{The time evolution of the SFE($R$) gradient as determined by linear fit to the bins from Figure \ref{f:hist3_Rsfe} plotted with $t^\prime$ time. For these fits, the prevalent central bin containing contributions from the intensely star forming nucleus region is excluded. The calculated gradient is marked by a blue square for IsoB and an orange triangle for TideB. A positive value indicates the overall SFE increases with radius while a negative value indicates a decreasing SFE with radius.}
    \label{f:hist_GradRsfe}
\end{figure}

To further understand any trends in this evolution of the radial dependence of star formation efficiency, we consider the change of the gradient in Figure \ref{f:hist3_Rsfe} with time for each case. To determine this gradient a linear least-squares fit was applied for all histogram bins excluding the centre-most bin defining $R < 1$\,kpc. The resulting first coefficient of this fit is plotted in Figure \ref{f:hist_GradRsfe} showing the gradient change with time. A positive result indicates the SFE increases overall with radius and, conversely, a negative result shows a decreasing SFE with radius. In this figure blue squares correspond to the calculated IsoB gradients, whereas TideB is denoted by orange triangles. It is clear that there is significant variation in the SFE  dependence with radius over the evolution time for both cases. During the first 50-100\,Myr after bar formation, the trend for each case appears similar: a steep negative gradient steadily begins to flatten and oscillate around zero, indicating average SFE across the disc is generally constant at that time. However, while the isolated case seems to remain mostly within $\pm 1$\,Myr$^{-1}$kpc$^{-1}$ of zero, the TideB results continue to climb, transitioning through zero to develop a clearly positive gradient for the duration of the second 100\,Myr and thus a decreasing depletion time increasing with radius. In the last 50\,Myr of the period considered, the TideB gradient does reduce to values possibly comparable to the steepest positive extent of the isolated case.There is one period of the IsoB case in later times which seems noticeably inconsistent with the other values. At $t^\prime = 160$\,Myr the gradient is almost the steepest negative result of both cases after bar formation. It is unclear from this analysis what kind of event caused such a strongly declining profile of SFE just for this brief time. However, it can be seen clearly in Figure \ref{f:hist_SplineXsfe} that there is a uncommonly high value of SFE across the length of the bar for this particular time period ($t^\prime = 160$\,Myr). Although, this is simply an illustration of  the mathematical cause for such a strongly declining trend and does not necessarily explain the physical origin. More discussion of the spatial and temporal variations of bar SFE is included in the following section. 

\begin{figure*}
	\includegraphics[width=\textwidth]{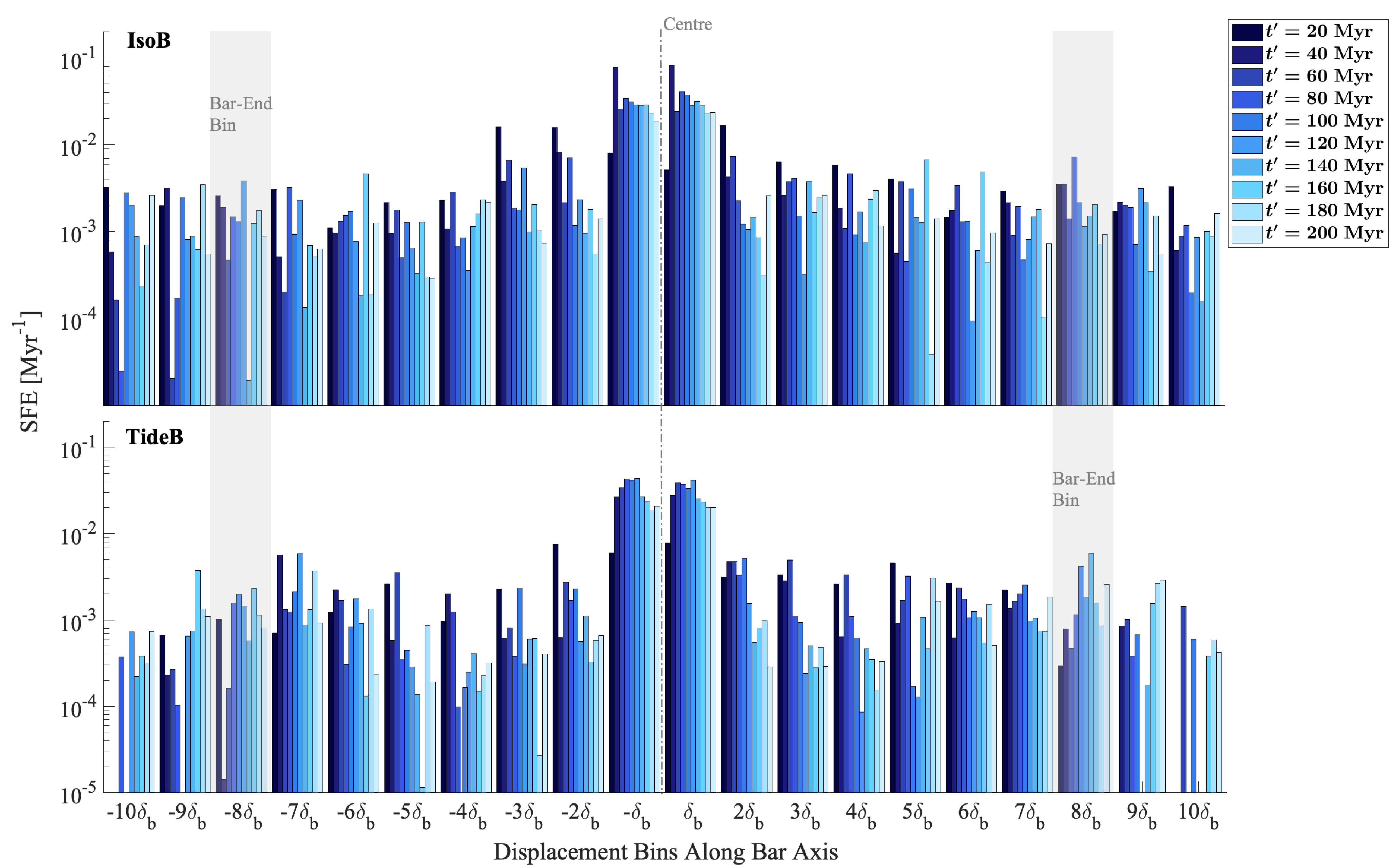}
    \caption{For each case, the SFE across the bar binning along the bar axis for the barred period ($t^\prime >  0$). The bar axis defined by a co-rotating co-ordinate system to always align with the $x$-axis. The bar thickness in the disc plane is set to be $y \pm1$\,kpc. As the bar length differs for each case, the bar is divided into a consistent number of bins so each bin corresponds to a fractional bar element ($\delta_b$). These are numbered outwards from the centre making it possible to simply assess symmetry between equidistant bar elements. The star formation time remains consistent with previous analysis ($\Delta t_{\rm SF} = 10$\,Myr). The time evolution of SFE can be seen with individually coloured bars corresponding to the result at time steps of 20\,Myr, with evolution from dark to light. The bin containing the bar extent is shaded in grey.  }
    \label{f:hist3_Xsfe}
\end{figure*}

In the context of observational studies, the results in Figure \ref{f:hist3_Rsfe} and Figure \ref{f:hist_GradRsfe} are generally consistent with observed SFE trends in galaxies, however the TideB response is somewhat significant. \citet{Leroy2008} measure the radial changes of SFE in the THINGS dataset \citep{Walter2008} and observe that most systems show that SFE in most galaxies tends to either maintain a constant value or decay with increasing radius. Some systems however, such as NGC 5194 (M51) and NGC3627 (the observational target for TideB), do conversely display localised regions with increasing SFE with radius in the disc \citep{Leroy2008}. This result is confirmed by \citet{Muraoka2019} with the COMING  \citep{Sorai2019} observational survey data showing that, while NGC3627 appears to have a generally constant SFE across the radii measured, there is clearly a slightly increasing trend at outer radii. M51 was not included in \citet{Muraoka2019} but the isolated target (NGC4303) was. Comparatively, NGC4303 shows a more consistently flat profile with smaller scale radial fluctuations in SFE across the disc \citep{Muraoka2019}. This observed SFE behaviour for NGC4303 is indeed similar to the IsoB result reflected in Figure \ref{f:hist_GradRsfe} which is primarily constant with only slight variation in both directions over time. M51 is a clearly interacting galaxy \citep{Leroy2008,Buta2019,Colbert2004,Karachentsev2013}; NGC3627 was specifically chosen for this study due to the possibility of an interaction rich history; and, the TideB response can also be seen to decay on a time-scale similar to the interaction time-scale. This may suggest that the characteristic of a positive SFE gradient toward outer radii is indicative of an ongoing, or at least recent, interaction. If this is the case, it would mean that an increasing SFE with radius may be used as a possible metric of tidally driven disc features. The relatively constant or negative gradients—which dominate the IsoB result and other times of TideB—can then also be considered compatible with more commonly observed trends, as instances where this may be observed for either IsoB or TideB over the duration of the simulation would indeed make it most common. 

\subsubsection{Dependence Along the Bar}
To specifically highlight changes within the bar, Figure \ref{f:hist3_Xsfe} shows a similar histogram representation of SFE to Figure \ref{f:hist3_Rsfe} this time binned along the bar axis. Many studies have shown that SFE in barred-spiral galaxies is lower in the bar than the associated arm features \citep{Momose2010,Hirota2014,Yajima2019}, and in some cases that SFE even varies across the length of the bar \citep{Downes1996,Sheth2002,Muraoka2019}. For the target galaxy NGC\,3627 specifically, it appears that there is unexpectedly higher SFE at the bar ends and centre than elsewhere \citep{Watanabe,Law2018}. Additionally, this work has shown in the previous sections that there are discernible differences in the shape of features within the bar region which can be seen in projections $\Sigma_{\rm SFR}$ across the disc (see Figure \ref{f:sfr_prj}).

\begin{figure*}
	\includegraphics[width=\textwidth]{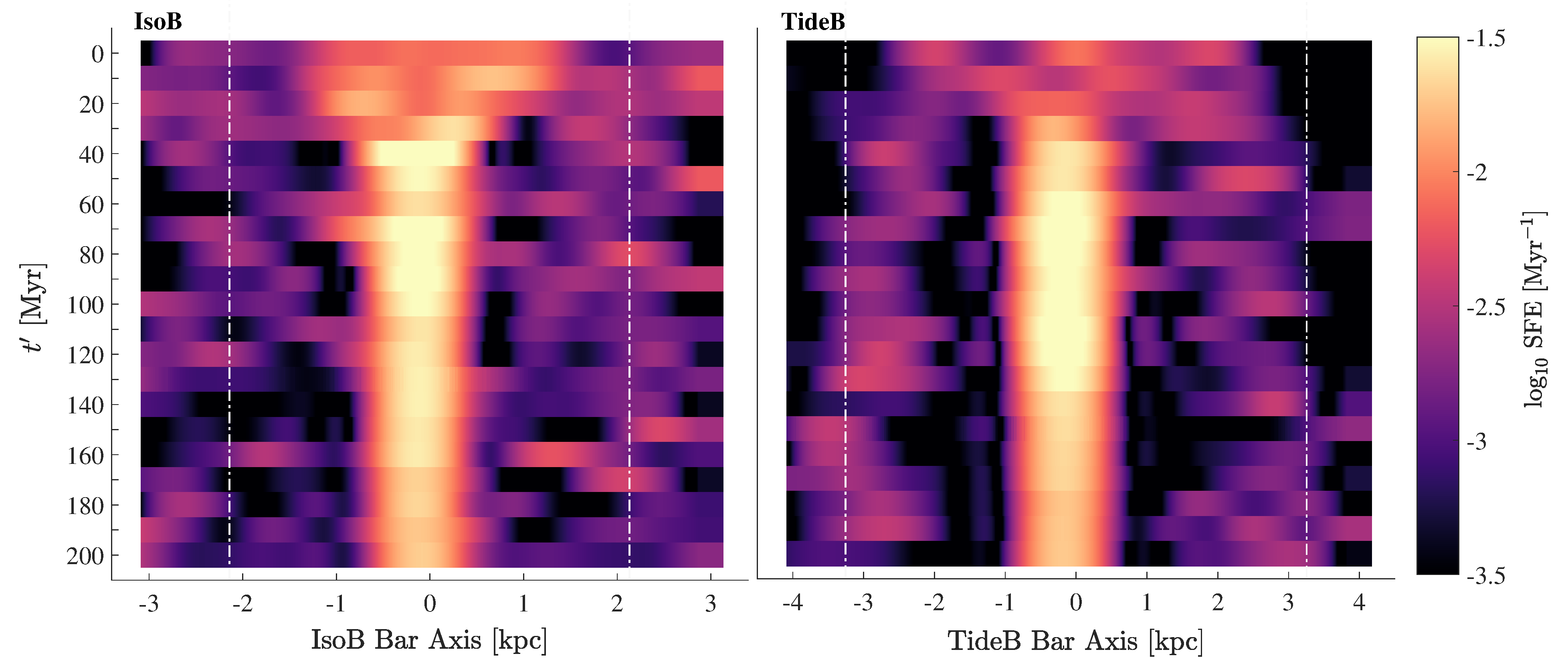}
    \caption{The time evolution of SFE along the bar axis from Figure \ref{f:hist3_Xsfe} is re-oriented as a heat-map for SFE smoothed by a spline interpolant. Each resolved $t^\prime$ corresponds to a strip of intensities along the $x$-axis which correspond to the SFE values in each bar element. Time evolution is downward along the $y$-axis. The colour weighting is such that brightest areas correspond to high efficiency and dark values to low efficiency. The bright central band corresponds to the nuclear disc, and the white dot-dashed lines illustrate the previously defined bar length.}
    \label{f:hist_SplineXsfe}
\end{figure*}

We define a bar axis and calculate the specific SFE in a polygon cut-out running parallel to this axis with a width of 2\,kpc. We consider the bar to be a solid body rotating about the major axis, as bar slow-down tends to occur on much longer time-scales than investigated here \citep{Miwa1998,Debattista2000,Kormendy2013}. The bar orientation is determined by identifying density peaks in the stellar population within the bar radius. The orientation of the bar axis is then used to transform all the simulated positional information to a co-rotating co-ordinate system wherein the bar lies along the $x-$axis. Additionally, the bar length varies between the two cases so, in order to produce comparable bin limits, the total bar length was divided into a set number of bins which would correspond to a fraction of the whole bar. These are labelled as dimensionless fractional bar elements ($\delta_b$) and numbered outwards from the centre-most bin to the calculated bar extent, plus two extra bins to encompass the bar end regions. While this indicates that the physical size of bins is different for each result, this difference is sufficiently negligible to the effect of scale on SFE results in this case. As in Figure \ref{f:hist3_Rsfe}, the star formation time is $\Delta t_{\rm SF} = 10$\,Myr and the time evolution is mapped by coloured bars corresponding to time steps of 20\,Myr evolving from dark to light. Similarly, the bin coinciding with the bar end is shaded in grey.  

Along the bar, we are looking to assess two key features: symmetry about $x = 0$ to consider whether star formation is symmetrical along the bar; and whether there is any noticeable trend in the SFE($x$) profile such as significant peaks or troughs in SFE at key regions (i.e. at the bar ends). Interestingly, in both cases the SFE profile along the bar does not seem particularly symmetrical at any of the time periods shown in Figure \ref{f:hist3_Xsfe}. However, neither does it seem consistently asymmetric or skewed to one particular direction at all times for either case.

Within each bin, there is also no standard trend of SFE changing with time, similar to the radial dependence. There is, however, noticeable evolution in the shape of the bar-wide trends in both cases. In the case of IsoB, at earlier times the SFE seems to decrease steeply toward the bar ends ($t^\prime < 100$\,Myr). In the period after $t^\prime = 100$\,Myr, the SFE values for each bin appear variable along the bar axis in the small scale but are generally constant on average. Comparatively, the TideB result also seems to predominantly show SFE decreasing steeply towards the bar end for the earliest times. After $t^\prime\sim 60$\,Myr, however, the SFE appears to drop in the middle bar ($2\delta_\mathrm{b} \lesssim |x| \lesssim 5 \delta_\mathrm{b}$) before rising again towards the bar edges. This is consistent with the previous section where a significant change in the SFE gradient with radius of the TideB case was observed at almost exactly this time period, showing lower SFE in the inner region of the disc compared to outer radii. 

To further quantify this variation of SFE along the bar, we re-orient Figure \ref{f:hist3_Xsfe} as a heat-map of the SFE along the bar axis in Figure \ref{f:hist_SplineXsfe}. In this figure, the SFE values for each time are smoothed by fitting a spline interpolant and then plotted along a strip of bar axis values in\,kpc for each case. Time evolution is defined as downward along the $y$-axis and the calculated result for each 10\,Myr. 
While no obvious trend appears outside the peak of the centre-most nucleus in the IsoB case, the bar in the TideB case quickly evolves to show strong SFE at the bar ends (delineated by white vertical dashed lines) and low SFE at approximately half the bar radius, alongside the expected high peak in the central region. The IsoB case does show a similar SFE deficiency at some time periods but this appears more serendipitous as a part of random variation which fluctuates along the entire length of the bar. From this figure, it is also evident that there is some variation in the intensity of SFE in the central region with time. This trend is similar between the two cases, appearing suddenly and showing brightest in the early period of the bar before decreasing slightly but steadily as the bars evolve.

\section{Discussion}
\label{s:discussion}
Studies of the interacting target (NGC 3627) specifically note observations show higher SFE at the bar ends and centre compared to other parts of the bar \citep{Watanabe,Law2018}. In comparison, studies of the isolated target (NGC4303) more commonly report more average values for SFE along the bar \citep{Momose2010}, although \citet{Yajima2019} also specifically investigate the bar end to also find higher results than the central bar or arm regions studied. In Figure \ref{f:hist_SplineXsfe} it can be seen clearly that while there is certainly some variation along the axis of a bar formed in isolation, it is much more likely to appear as featureless on average. In this case, resolution and the specific locations and shapes used to classify features could significantly impact the overall results observed. Comparatively, there must be some active effect in the TideB case which is obviously causing star formation to be consistently strongest at the bar edges and centre but severely suppressed in the regions between these areas. 

Variation in SFE within the bar region has previously been attributed to possible non-circular motions generating strong shocks or shear motions along the bar and disrupting many bar-located molecular clouds \citep{Schinnerer2002,Dobbs2014}. It is true that the TideB case certainly shows significantly higher non-axisymmetric motion (see Figure \ref{f:velocity}), however these stronger motions are also evident across the entire disc. \citet{Law2018} draw a possible correlation between kinetic temperature and SFE in NGC 3627 (TideB target), which may support this. 

\begin{figure}
	\includegraphics[width=\columnwidth]{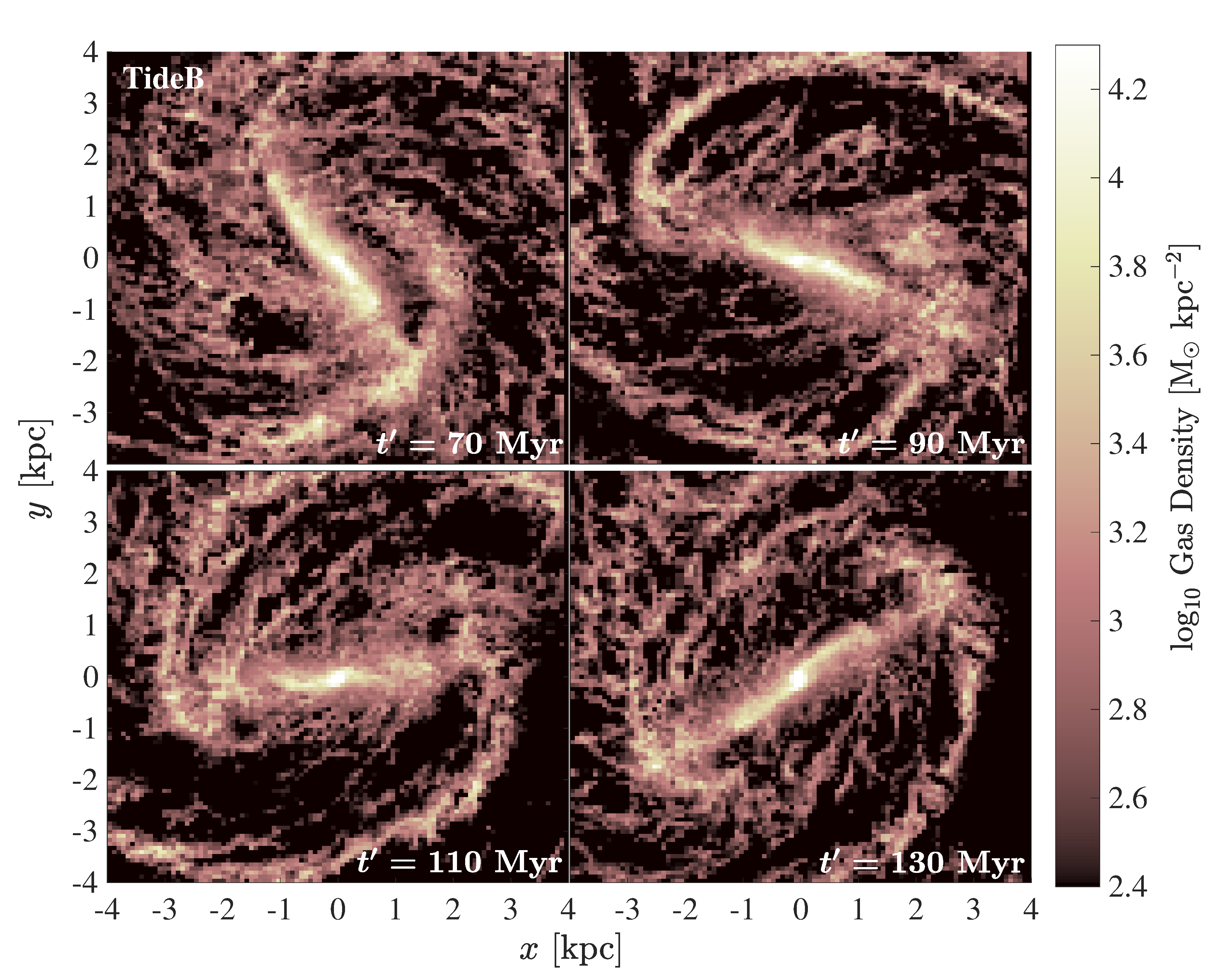}
    \caption{The TideB central zoom-in to highlight the bar-spiral overlap regions. Two time steps equally spaced either side of $t^\prime = 100$\,Myr. }
    \label{f:zoom_decoupling}
\end{figure}

Contrary to the above, \citet{Beuther2018} assert that the surface densities of NGC 3627 are too high for shear to be sufficient to effectively dampen star formation activity at the bar ends, and that differing pattern speeds between the bar and arms may be the most favourable condition to promote intense star formation in these regions. The arms in the TideB model do frequently decouple from the bar ends and reconnect, significantly more so than the IsoB case. Figure \ref{f:zoom_decoupling} shows a magnified view of the TideB gas in the central region for two time periods either side of the $t^\prime = 100$\,Myr result shown in previous sections. The arch-like arm features bracket the bar ends, highlighting their decoupling from the bar; a time-dependent process which appears to occur frequently over orbital time-scales. Calculating the pattern speed for both the bar and arms it is possible to determine that there is a difference in rotation speed at radii about the bar end. The bar pattern speed was determined by tracking the angular displacement of the defined bar axis at the given radius $R_{\rm{bar}}$ over the duration of interest $t^\prime>0$ which is in line with common methods considered appropriate for resolving transient features \citep{Grand2013,Pettitt2017}. Throughout the relatively short period of interest the bar pattern speed does not significantly change from a value of $\Omega_{\rm{bar}}=25.9$\,km\,s$^{-1}$\,kpc$^{-1}$ for the TideB case. Considering the arm pattern speed to be dynamic, transient and tidally driven, we assume an arm pattern speed of $\Omega_{\rm{sp}} = \Omega({\rm disc}) - \kappa/2$ in line with \citet{Pettitt2016}. At the bar end $R_{\rm{bar}}$, this radially dependent pattern speed is found to be $\Omega_{\rm{sp}}=41.9$\,km\,s$^{-1}$\,kpc$^{-1}$ which is significantly greater than the bar speed and thus, evidence of such decoupling between the two features.

Considering the continued influence of the companion, it is also possible that the sustained torques from the interaction may impact the development of such an obviously high SFE at the bar ends. It is possible that the initial starburst effect triggered by the interaction may be responsible. Under this kind of driven bar formation, the interaction generates a large inflow toward the centre and causing increased star formation in the entire central region. However, over time the available matter may become insufficient to support the same amount of star formation after the initial in-flow amount is converted. This may be in line with observations of NGC\,3627 where \citet{Casasola2011} postulate that the bar end may be aligned with co-rotation and that the absence of an inner Lindblad resonance (ILR) means the torques are negative between the bar and the nucleus. In this state, NGC\,3627 can be considered a so-called \textquoteleft smoking gun' of inner gas inflow, wherein the dynamical resonances and kinematically decoupled inner bar serve to directly supply gas to the central region and AGN \citep{Casasola2011}. Although \citet{Casasola2011} note that this process should be unsustainable after the bar has had time to slow under secular evolution, the bar pattern speed of TideB does not significantly slow over the relatively short duration of this analysis period and so discernible traces should still be evident. However, calculating the co-rotation radius for the TideB disc gives a radius $R_{\rm{CR}}=4.97$\,kpc which is not especially close to the determined bar extent ($R_{\rm{bar}}=3.12$\,kpc). So, dynamical resonances directly supplying the inner region with gas in this way is unlikely. Additionally, such an explanation does not necessarily account for why there should remain only preferentially strong star formation at the bar ends, in addition to the centre.

These are all possible factors which may impact the formation of such a consistent feature in tidally driven bars which does not appear to persist for any length of time in the more isolated evolutionary scenario. This appears to indicate there is a need for further, more directed studies that will facilitate a determination of the root cause and relative probability of such a feature occurring in different galaxy populations, as well as how such features may propagate into the later stages of bar and disc evolution. This is particularly pertinent as these clear fissures in the SFE profile are both prominent and enduring in the early stages of the tidal bar, so would make a fitting metric by which observational surveys may determine the formation mechanism of a given galactic bar.

\section{Conclusions}
\label{s:conc}
The primary aim of this work was to consider the effects that the formation mechanism driving a spiral galaxy to form a bar may have on the resulting properties of the host galaxy, in particular on its star forming potential. To achieve this, two hydrodynamical $N$-body simulations were performed: IsoB which would form a bar by isolated, independent evolution; and TideB which required the impetus of a passing companion to trigger bar formation through tidal interaction. These two scenarios were designed to mimic the nearby spirals NGC\,4303 (IsoB) and NGC\,3627 (TideB), popular barred-spiral targets for observational surveys. These simulations produced two clearly barred-spiral galaxies with similar morphologies to the real target galaxies. Both cases show the evolution of similar exponential type bars with a two-arm, grand design spiral-like structure in the stellar and gas components. Velocity profiles in the gas disc of both cases are also relatively similar in terms of the shape of features across the disc. The degree of non-axisymmetric motion in the tidally driven TideB case is, however, significantly greater than that of the isolated IsoB case at all evolutionary time periods.

The star formation history and evolution of features in a SFR projection onto the disc plane shows that both cases have relatively similar levels for average SFR once the bar is formed, including a prolonged burst of bar-driven star formation. However, in the TideB case there is also an additional burst-like period of star formation which is likely caused by the interaction strengthening the arms before the bar forms. This boost is conincident with a period where significant star formation is apparent in the arm regions of TideB that is not seen in the history of the IsoB case. Both cases also show clearly distinguishable star forming structures within the bar radius which differ independently with time, as well as between each formation mechanism. The general shape of the $\Sigma_{\rm gas}$ vs. $\Sigma_{\rm SFR}$ distribution clearly varies depending on the bar formation mechanism. The TideB case shows a much higher distribution of inefficient but high gas density regions, though these seem to be confined to tidal debris in the outer disc. However, both cases appear to conform to the literature Kennicutt-Schmidt relation averaged over all galaxies. 

While many general properties of each case appear similar, there are in fact, significant variations in star formation properties within each of the bar, arm or even inter-arm regions. Considering the Kennicutt-Schmidt relation in terms of different morphological components (bar, arm and inter-arm) shows that there is little visually distinguishable difference between the arm and inter-arm regions except for the arms being further along the KS relation than the inter-arm on average, and the bar further along than both. The bar region in all cases follows a significantly steeper profile than any other component and the disc as a whole. The arm region is comparatively similar to the full disc average for both the isolated and tidal bars. The inter-arm region represents the shallowest profile in all cases, comparable only to the disc average slope if tidal remnants in the far outer radii are included in the TideB case.

The overall radial dependence of SFE in the early stages of the bar evolution in both cases shows a strongly decreasing profile with increasing radius. For the IsoB case, this profile quickly flattens once outside of the very central region ($R > 1$\,kpc). However, the TideB case eventually reverses the gradient of SFE across the disc, showing a steady positive gradient with SFE, increasing towards outer radii (excluding the centre-most region) at later times. This is attributed to the apparent development of a region of suppressed star formation in the middle extent of the bar in the TideB case. These features are likely related to the presence of the companion in some way, possibly due to shocks or shearing effects from the strong non-axisymmetric motions present, tidal torques or the result of the initial burst quickly absorbing available resources in that region. However, it remains that the tidally driven case shows a particularly consistent and identifiable SFE profile along the bar axis with peaks at the bar ends and in the centre, not seen in the isolated bar. Further investigation into the presence, propagation and drivers of this SFE deficiency is identified as worthwhile as it may allow for a consistent identification method for bar formation mechanism in observable galaxies, since it is only evident in the tidally driven bar. 

\section*{Acknowledgements}
We thank the anonymous referee whose comments helped improved the quality of this manuscript.
EJI acknowledges the support of Japanese Government MEXT Scholarship for Foreign Students.
ARP acknowledges the support of The Japanese Society for the Promotion of Science (JSPS) KAKENHI grant for Early Career Scientists (20K14456). 
TO acknowledges the support of MEXT KAKENHI grants 18H04333, 19H01931, and 20H05861. 
Images and partial analysis were made using the \textsc{pynbody} \textsc{python} package (https://github.com/pynbody/pynbody, \citep{Pontzen2012} and the MathWorks \textsc{matlab} programming and numeric computing platform. Observational data was acessed from the PHANGS-HST survey \citep{Lee2021}. Numerical computations were carried out on the Cray XC50 at Center for Computational Astrophysics, National Astronomical Observatory of Japan. We thank K. Sorai, Y. Yajima and S. Benincasa for helpful discussions realated to this work.

\section*{Data Availability}
The data underlying this article can be shared upon reasonable request to the corresponding author. 



\bibliographystyle{mnras}
\bibliography{bib_thesis} 




\appendix

\section{Rotation Curve Evolution}
\label{a:rc}
\begin{figure}
	\includegraphics[width=\columnwidth]{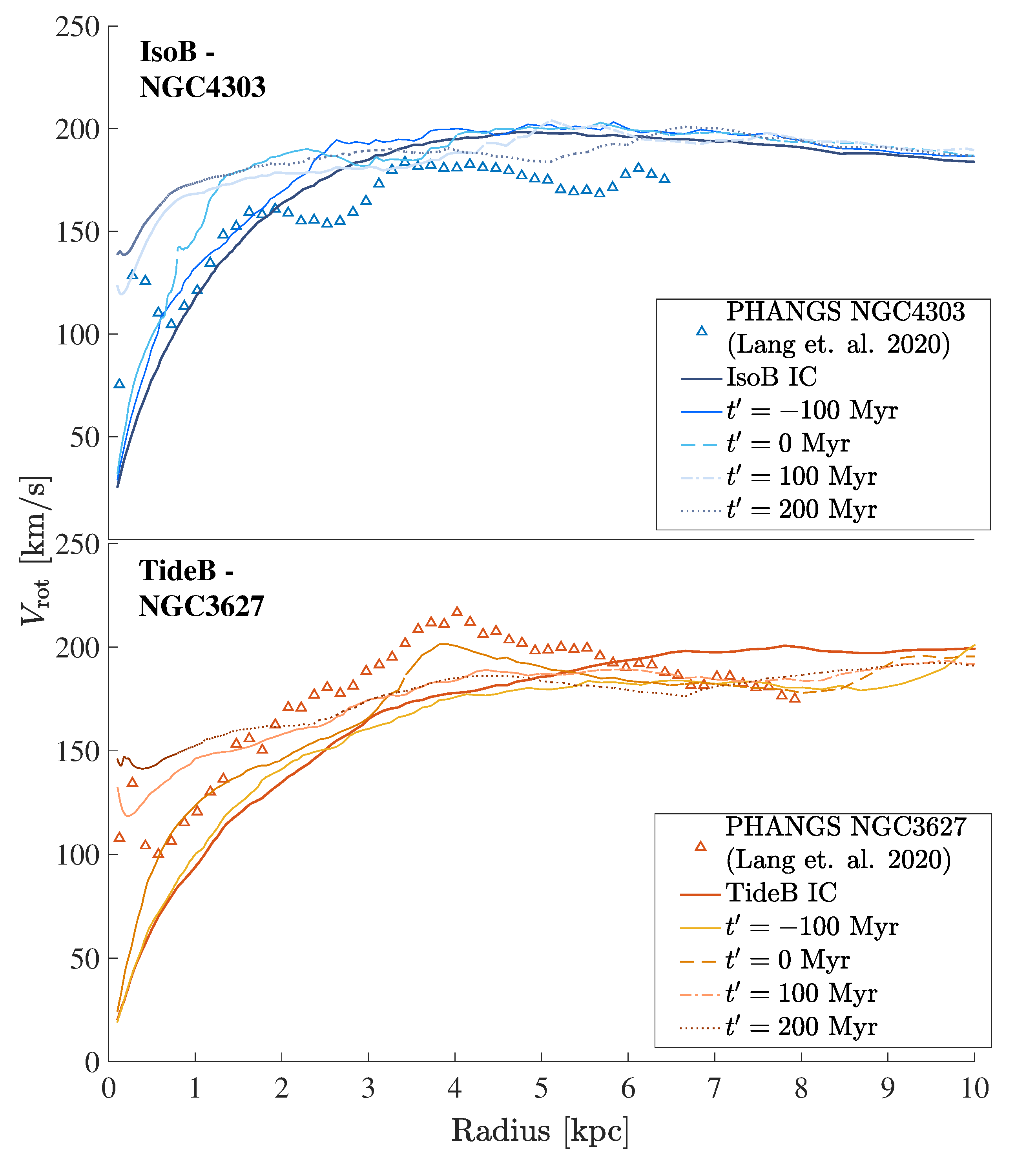}
    \caption{Rotation curves for each case at the key intervals $t^\prime=[-100, 0, 100, 200]$\,Myr as well as the initial condition and observation results from the PHANGS survey data \citep{Lang2020}.}
    \label{f:RC_evolution}
\end{figure}

In Figure\;\ref{f:RC_evolution} we show the rotation curve in both models (IsoB top, TideB bottom) at different evolutionary stages. There is clearly a significant variance in the shape of the rotation curve over the time period investigated. Both curves tend to rise up in the inner disc as the bar grows, with outer undulations decaying on 100\,Myr time-scales. Of particular note is the spike around 4\,kpc in the TideB case that lines up particularly well with the NGC\,3627 profile at $t'=0$\,Myr, which is also coincidentally when the model shows the closest morphological match.

\bsp	
\label{lastpage}
\end{document}